\documentclass[final,5p,times,twocolumn]{elsarticle}

\usepackage{url}

\usepackage{graphicx}
\graphicspath{{./figures/}}
\DeclareGraphicsExtensions{.png}
\usepackage{subfig}

\usepackage{graphicx}

\usepackage{amssymb}
\usepackage{amsthm}

\usepackage[T1]{fontenc}
\usepackage{textcomp}
\usepackage{lmodern}

\journal{Preprint}

\begin{document}

\begin{frontmatter}



\title{PhoenixCloud: Provisioning Resources for Heterogeneous Workloads in Cloud Computing}

 \author[label1]{Jianfeng Zhan\corref{cor1}\fnref{fn1}}
 \cortext[cor1]{Corresponding author}
 \fntext[fn1]{Tel:010-62601006;}
 \fntext[fn2]{This is an extended version of our CCA 08 paper(The First Workshop of Cloud Computing and its Application, CCA08, Chicago, 2008): J. Zhan L. Wang, B. Tu, Y. Li, P. Wang, W. Zhou, D. Meng. 2008. Phoenix Cloud: Consolidating Different Computing Loads on Shared Cluster System for Large Organization. The modified version can be found on http://arxiv.org/abs/0906.1346.}
\ead{jfzhan@ncic.ac.cn}
\author[label1]{Lei Wang}
\ead{wl@ncic.ac.cn}
\author[label2]{Weisong Shi}
\ead{weisong@wayne.edu}
\ead{jfzhan@ncic.ac.cn}

\author[label1]{Shimin Gong}
\ead{gongshimin@ncic.ac.cn}

\author[label1]{Xiutao Zang}
\ead{zangxiutao@ncic.ac.cn}

\address[label1]{Institute of Computing Technology, Chinese Academy of Sciences, Beijing 100190, China}
\address[label2]{Department of Computer Science, Wayne State University,Detroit, MI 48084, USA}

\begin{abstract}
As more and more service providers choose Cloud platforms, which is provided by third party resource providers, resource providers needs to provision resources for heterogeneous workloads in different Cloud scenarios. Taking into account the dramatic differences of heterogeneous workloads, can we coordinately provision resources for heterogeneous workloads in Cloud computing? In this paper we focus on this important issue, which is investigated by few previous work. Our contributions are threefold: (1) we respectively propose a coordinated resource provisioning solution for heterogeneous workloads in two typical Cloud scenarios: first, a large organization operates a private Cloud for two heterogeneous workloads; second, a large organization or two service providers running heterogeneous workloads revert to a public Cloud; (2) we build an agile  system \emph{PhoenixCloud} that enables a resource provider to create coordinated runtime environments on demand for heterogeneous workloads when they are consolidated on a Cloud site; and (3) A comprehensive evaluation has been performed in experiments. For two typical heterogeneous workload traces: parallel batch jobs and Web services, our experiments show that: a) in a private Cloud scenario, when the throughput is almost same like that of a dedicated cluster system, our solution decreases the configuration size of a cluster by about 40\%; b) in a public Cloud scenario, our solution decreases not only the total resource consumption, but also the peak resource consumption maximally to 31\% with respect to that of \emph{EC2} +\emph{RightScale} solution. 
\end{abstract}


\end{frontmatter}


\section{Introduction}
Traditionally, users tend to use a dedicated cluster system (in short DCS) to provide homogeneous services. Resource utilization rates of DCS are varying. For unexpected peak loads, DCS cannot provision enough resources, while lots of resources are idle for normal loads. Recently, several pioneer computing companies are adopting infrastructure as a service (IaaS). For example, as a resource provider, Amazon provides elastic computing cloud (EC2) services \cite{8} to end users in order to offer outsourced resources in the granularity of XEN virtual machine \cite{39}. A new term Cloud is used to describe this new computing paradigm \cite{5} \cite{33} \cite{41}. We regard that the most appropriate one is defined in \cite{38}. According to this definition, \emph{a Cloud is a large pool of easily usable and accessible virtualized resources, which can be dynamically reconfigured to adjust to a variable load (scale), allowing also for optimum resource utilization}.\\
As more and more service providers choose Cloud platforms, which is provided by third party resource providers, a resource provider (which can be regarded as a \emph{Cloud infrastructure provider}) needs to provision resources for heterogeneous workloads in Cloud computing. In this paper, we consider two representative Cloud scenario: first, a large organization operates a private Cloud for heterogeneous workloads; second, a large organization or service providers running heterogeneous workloads revert to a public Cloud. For example, a large organization operates two DCSs for its two affiliated departments: a batch queuing system for parallel batch jobs for the first department and a Web service infrastructure for the second one. \emph{Is it possible for this organization to resort to a private Cloud or a public Cloud solution}? Besides, independent service providers also may run different heterogeneous workloads. Heterogeneous workloads have different resource management requirements, for example workloads of parallel batch jobs and Web services differ in resource consumption characteristics, performance goals and time scales of management\cite{37}, (which we will further explain in Section \ref{requirement}). \emph{Taking into account the dramatic differences of heterogeneous workloads, can we coordinately provision resources for heterogeneous Cloud workloads?}  This issue is the main focus of our paper. Besides, \emph{the runtime environment software that is responsible for managing cluster resources and workloads} plays an important role since it has great impact on resource utilizations and quality of services of user applications. Traditional runtime environments only support homogeneous workloads, for example, \emph{OpenPBS} \cite{30} for parallel batch jobs or \emph{Oceano} \cite{1} for web services. And hence we also need to consider another supporting issue in this paper:\emph{ how should we design and implement runtime environment software that enables provisioning resources for heterogeneous workloads in different Cloud scenario?}\\
Previous work fails to resolve the above issues in two ways:
First, previous work either devises a scheduling algorithm for parallel batch jobs with different resource demands\cite{35}, either proposes resource allocation algorithms for virtualized service hosting
platforms \cite{47}, of which clustered servers run components of \emph{continuously running services}, or presents resource management policies (implementing leases as virtual machines \cite{36}) for homogeneous workloads (only parallel batch jobs) mixed with \emph{best-effort lease requests and advanced reservation requests} in Cloud scenarios, and hence we can not leverage existing knowledge to resolve coordinated resource provisioning issues for heterogeneous Cloud workloads. \\ Second, though previous systems can provision virtual infrastructure \cite{45} \cite{46} or hosted application environments \cite{12} in private/hybrid clouds, no previous efforts pay attention to emerging requirements for coordinated resource provisioning for heterogeneous workloads, and no system enables creating coordinated runtime environments on demand. \emph{A coordinated runtime environment is the one that can share coordinated resources with another runtime environment}. For example, if the large organization chooses a Cloud platform, two runtime environments belong to this condition.
To the best of our knowledge, this is the first time that the above issue is focused on. We design and implement an innovative system, PhoenixCloud, to facilitate a resource provider to provision coordinated runtime environments on demand for heterogeneous workloads in Cloud computing. The contributions of our paper are concluded as follows:\\
(1)In two typical Cloud scenarios, we respectively propose a coordinated resource provisioning solution for two representative heterogeneous workloads (parallel batch jobs and Web services): first, a large organization operates a private Cloud for two heterogeneous workloads; second, a large organization or two service providers running heterogeneous workloads revert to a public Cloud.\\
(2)We build an innovative system \emph{PhoenixCloud} to enable creating coordinated runtime environments for heterogeneous workloads.\\
(3) A comprehensive evaluation has been performed in experiments. For typical workload traces of parallel batch jobs and Web services, our experiments show that: a) in the private Cloud scenario, when the throughput is almost same like that of DCS, our solution decreases the configuration size of cluster by about 40\%; b) in the second Cloud scenario, our solution decreases not only the total resource consumption, but also the peak resource consumption maximally to 31\% with respect to that of EC2 + RightScale solution.\\
This paper includes seven sections. Section \ref{related_work} summarizes the related work. Section \ref{requirement} introduces several representative runtime environment requirements. Section \ref{design} explains PhoenixCloud design and implementation. Section \ref{policy} proposes two policies for coordinated resource provisioning. Section \ref{evaluation} evaluates our system, and Section \ref{conclusion} draws the conclusion.

\section{Related Work} \label{related_work}
In this section, we summarize related work of \emph{agile infrastructure} and \emph{resource provisioning}.

\subsection{Agile Infrastructure: Description Models and Systems}

\textbf{Description models}: EC2 allows end users to describe their resource requirements, e.g., virtual machines, and EC2's extended services - RightScale \cite{32} allow service providers to describe their Web service requirements; A. Keller et al \cite{25} propose a framework to specify service-level agreements for web services; A. Hoheisel et al \cite{15} present a framework to define both workflow and dataflow for job applications. F. Gal¨¢n et al \cite{12} propose a service specification language for cloud computing platforms in order to facilitate interoperability among IaaS clouds, and also address important issues such as custom automatic elasticity and performance monitoring. R. Buyya et al\cite{38} propose a meta-negotiation document to determine definition and measurement of user QoS parameters. However, they are not qualified for describing diverse runtime environment requirements in creating coordinated runtime environments on demand.\\
\textbf{Systems}: EC2 directly provisions resources to end users. Without enabling the user role of service provider, EC2 relies upon end user's manual management of resources. EC2 extended services: RightScale \cite{32}, Enomalism \cite{9} and GoGrid \cite{13} systems provide automated cloud computing management systems that assist you in creating and deploying \emph{only scalable Web service applications} running on EC2 platforms. D. Irwin et al \cite{6} share a similar goal of our work by providing  Shirako prototype of a service oriented architecture for resource providers and consumers to negotiate access to resources over time; however Shirako does not explicitly support service providers to express personalized runtime environment requirements, especially coordinated runtime environments for heterogeneous workloads.\\
M. Steinder et al \cite{37} show that a virtual machine allows heterogeneous workloads to be collocated on any server machine, and proposes a system architecture for managing heterogeneous workload. However, it does not treat runtime environment as a first-class entity in the design. In our opinion, \emph{runtime environment's being a first class entity} has three meanings: a) there is a runtime environment specification that is qualified for expressing diverse runtime environment requirements; b) a runtime environment can be created on demand according to a runtime environment agreement; c) there is a framework that supports the development of a a runtime environment satisfying the new requirement. R. S. Montero et al \cite{29} propose an architecture to provision computing elements that focuses on resolving the growing heterogeneity (hardware and software configuration) of the organizations that join a Grid. A. Bavier et al \cite{4} demonstrate dynamic instantiation of distributed virtualization in a wide-area testbed deployment with a sizable user base, whereby each service runs in an isolated slice of PlanetLab's global resources.\\
B. Rochwerger et al \cite{27} pay attention to implementing an architecture that would enable providers of cloud infrastructure to dynamically partner with each other. Two open source projects, OpenNebula (\url{http://www.opennebula.org/}) and Haizea (\url{http://haizea.cs.uchicago.edu/}), are complementary and can be used to manage Virtual infrastructures in private/hybrid clouds [45][46]. However, those systems do not enable creating coordinated runtime environments for heterogeneous Cloud workloads.

\subsection{Resource provisioning}

M. Steinder et al \cite{37} only exploits a range of new automation mechanisms that will benefit a system with a homogeneous, particularly non-interactive workload by allowing more effective scheduling of jobs. By considering a workload in which massively parallel tasks that require large resources pools are interleaved with short tasks that require fast response but consume fewer resources, M. Silberstein et al \cite{35} devise a scheduling algorithm. In nature, they only consider parallel batch jobs with different resource demands. M.W. Margo et al \cite{28} are interested in metascheduling capabilities (co-scheduling for Grid applications) in the TeraGrid system, including user-settable reservations among distributed cluster sites.
B. Lin et al provide an OS-level scheduling mechanism, \emph{VSched} \cite{26}. VSched enforces compute rate and interactivity goals for interactive workloads, including web workloads and non-interactive ones. It provides soft real-time guarantees for VMs hosted \emph{on a single server machine}. \\
L. Grit et al \cite{14} design a Winks scheduler to support a weighted fair sharing model for a virtual "cloud" computing utility, such as Amazon's EC2, where each request is for a lease of some specified duration for one or more virtual machines. The goal of the Winks algorithm is to satisfy these requests from a resources pool in a way that preserves the fairness across flows, while our work focuses on how to provision resources for heterogeneous workloads when they are consolidated on a Cloud site. M. Stillwell et al \cite{47} proposes resource allocation algorithms for virtualized service hosting platforms, of which clustered servers run components of continuously running services, such as Web and e-commerce applications. In nature, they only consider homogeneous workloads.
B. Sotomayor et al \cite{36} present the design of a lease management architecture, which implements leases as virtual machines, to provide leased resources with customized application environments that only consider homogeneous workloads (only parallel batch jobs) mixed with best-effort lease requests and advanced reservation requests. VSched (\emph{as an OS-level scheduling mechanism}) or Haizea (\emph{for parallel batch jobs mixed with best-effort lease requests and advanced reservation requests}) can be used as a component of our system for specific workloads, when we consider to support more workloads.

\section{Diverse Runtime Environment Requirements}\label{requirement}
In this section, we summarize several representative cases of runtime environment requirements on a Cloud site.
\begin{itemize}
  \item Case One: Some universities are trying outsourcing of HPC services, just taking in this way the role of job-execution service providers \cite{3}.
  \item Case Two: many small companies have reverted to hosting environments for deploying Web services so as to decrease cost.
  \item Case Three: a large organization has two representative departments: a batch queuing system for parallel batch jobs for the first department and a Web service infrastructure for the second one. Instead of operating two DCSs, the organization either wants to consolidate heterogeneous workloads on a private Cloud or resorts to a public Cloud.
\end{itemize}

Three observations can be derived from the above three cases:
\begin{enumerate}
  \item There are three main user roles in the observed systems: \emph{resource providers}, \emph{service providers} and \emph{end users}. For example, in Case two, universities play the role of service providers, and they want to outsource resources to a resource provider and run batch queue systems for end users, including graduate students or researchers.
  \item A resource provider does need to provision resources and create runtime environments for heterogeneous workloads. For example, when the organization in Case Three chooses a private Cloud or resorts to a public Cloud, or two service providers in Case one and Case two resort to a public Cloud, a resource provider requires provisioning two different runtime environments for heterogeneous workloads.
  \item For heterogeneous workloads, runtime environment requirements are dramatically different. Coordinated resource provisioning for heterogeneous workloads may bring benefits to service providers and resource providers. \end{enumerate}

For example, runtime environments for parallel batch jobs and Web services differ in four ways:
\begin{itemize}
  \item Workloads are different. Web service workloads are often composed of a series of requests; while parallel batch job workloads are composed of a series of submitted jobs, and each job is a parallel or serial application.
  \item Resource consumptions are different. Running a parallel application needs a group of exclusive resources. While for Web services, requests will be serviced simultaneously and interleavedly through multiplex use of resources.
  \item Performance goals are different. From perspectives of end users, for parallel batch jobs, in general submitted jobs can be queued when resources are not available. However, for Web services like Web servers or search engines, each individual request needs an immediate response.
  \item Time scales of management are different \cite{37}. Due to the nature of their performance goals and short duration of individual requests, Web services need automation at short control cycles, e.g., seconds; However, parallel batch jobs typically require calculation of a schedule for an extended period of time \cite{37}, e.g., hours.
\end{itemize}
 	 	
When web service applications and parallel batch jobs are consolidated, we can propose coordinated resource provisioning solutions, since they have different performance goals.

\section{PhoenixCloud Design and Implementation } \label{design}

In Section \ref{objective}, we introduce the objectives of PhoenixCloud. Section \ref{specification} proposes a \emph{runtime environment specification}. In Section \ref{architecture}, we describe the architecture.
\subsection{Objectives} \label{objective}
PhoenixCloud has several objectives:
\begin{enumerate}
  \item Responsibility division between a resource provider and service providers. In our system, a resource provider is responsible for creating, destroying runtime environments and provisioning resources to different runtime environments on a Cloud site, while a service provider only focuses on providing service.
  \item Provisioning a runtime environment on a basis of a runtime environment specification. PhoenixCloud provides a runtime environment specification for a service provider to express runtime environment requirements. According to a runtime environment specification, a runtime environment is provisioned on demand.
  \item Pluggable resources type \cite{21}. Similar to \emph{Shirako}, provisioned resources will include servers, storages, and network resources. Presently, our system mainly facilitates provisioning servers in the granularity of node or virtual machine.
  \item Coordinated resource provisioning for heterogeneous workloads. If allowed by service providers, PhoenixCloud supports coordinated resource provisioning for two heterogeneous workloads.
\end{enumerate}

PhoenixCloud evolves from our previous Phoenix system \cite{40}. We have implemented PhoenixCloud on the Dawning 5000 cluster system, which is ranked as top 10 of Top 500 super computers in November, 2008. It is expected that PhoenixCloud will be deployed on the super computer-Dawning 6000 system in Shenzhen super computing center, China, in 2010.

\subsection{Runtime Environment Specification} \label{specification}
\begin{figure}[h]
\centering
\includegraphics[width=2.5in]{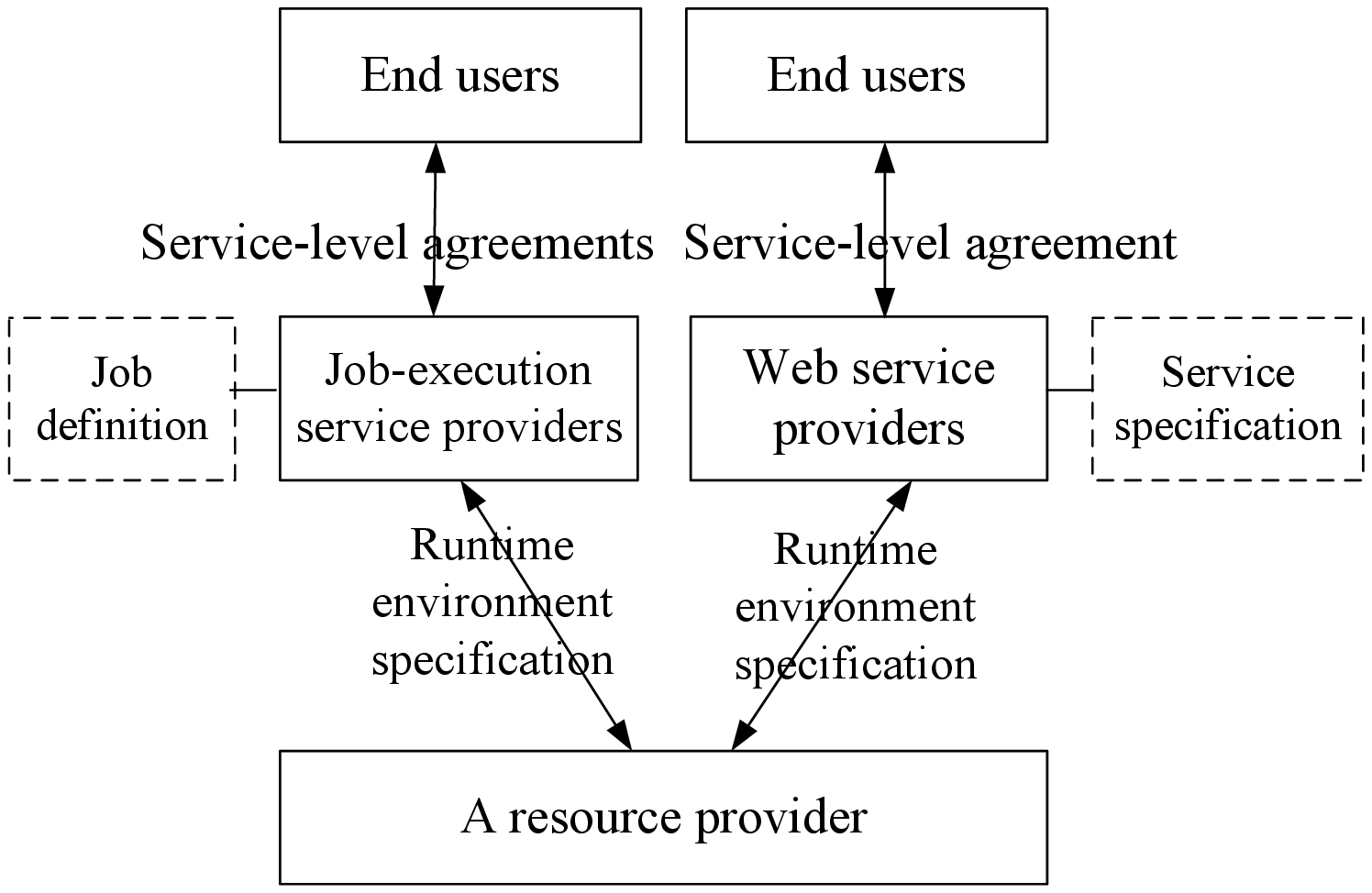}
\caption{The agreements among a resource provider, service providers and end users.}\label{relation_agreements}
\end{figure}

We present a runtime environment specification as a basis for provisioning a runtime environment or two coordinated runtime environments on demand.
In our opinion, in addition to service-level agreements between service providers and end users, job definitions for computational applications and service definitions for web services, both a resource provider and a service provider need a runtime environment specification to express diverse runtime environment requirements, on a basis of which, a resource provider can flexibly provision runtime environments on demand for service providers. Figure \ref{relation_agreements} shows the relationships of different agreements among different roles.

A runtime environment specification includes information as follows:
\begin{enumerate}
  \item Relationships between a service provider and a resource provider.\\
We support three different relationships: \emph{same} or \emph{affiliated} or \emph{business}. \emph{Same} means that a single user plays the roles of both resource provider and service provider, which describes traditional DCS; \emph{affiliated} means that a user playing the role of service provider is affiliated to a user playing the role of resource provider, which can describe Case Three in Section \ref{requirement}; \emph{business} means that a service provider has the business relationship with a resource provider, which can describe Case One or Case Two in Section \ref{requirement}.
  \item Workload types.\\
We supported two workloads types: parallel batch jobs and Web services. we are extending to support MapReduce and Dryad applications in Cloud.
  \item The allocation granularity of resources.\\
We support resource allocation in the granularity of nodes or virtual machines like XEN. For virtual machines, we provide predefined or user-defined virtual machine templates. For both nodes and virtual machines, users need to specify customized operating system types and versions.
  \item Coordinated runtime environments.\\
A service provider needs to decide two conditions: (a) whether a new runtime environment has a coordinated runtime environment that belongs to the same service provider; (b) Whether a service provider agrees that a new runtime environment is coordinated to share resources with other runtime environment of \emph{another service provider}.
  \item Resource coordination models and bound sizes of resources.\\
In each runtime environment, a service provider needs to specify two optional resource bounds: \emph{the lower resource bound} and \emph{the upper resource bound}. The lower resource bound is rigid in that a resource provider will guarantee that resources within the limit of the lower resource bound will only be allocated to a runtime environment or its coordinated runtime environment. The upper resource bound is flexible in that resources within the range defined by the lower resource bound and the upper resource bound, which firstly satisfy resource requests of the specified runtime environment or its coordinated runtime environments, can be reallocated to another runtime environment when they are idle. Fig.\ref{resource_bound} shows the relationship between two bound sizes of resources.\\
For two typical heterogeneous workloads: Web services and parallel batch jobs, we respectively propose a resource coordination model in two Cloud scenarios.\\
In the private Cloud scenario, we presume that a resource provider owns the fixed resources in a private Cloud that satisfy resource requests of two coordinated runtime environments. For this scenarios, we set the same size for both the lower resource bound and the upper resource bound. For two coordinated runtime environments, the size of the coordinated resources that are shared by two runtime environments is the sum of the lower resource bounds of two runtime environments. We call this model a FB (Fixed Bound) model.

In the public Cloud scenario, we presume that a resource provider owns enough resources that can satisfy resource requests of $N$ service providers ($N>>2$). For a runtime environment, we only specify the lower resource bound size with the upper resource bound size undefined.  Each runtime environment can request more resources beyond the limit of the lower resource bound.\\
For coordinated resource provisioning, we choose the following principles:

\begin{itemize}
  \item If a service provider does not allow coordinated resource provisioning, a runtime environment will be provisioned independently. In this case, RightScale \cite{32} and our previous effort \cite{41} respectively show individual Web service applications or scientific computing workloads like parallel batch jobs can benefit from the elastic management of Cloud.
  \item If allowed by service providers, we support coordinated resource provisioning at \emph{a granularity of a group, which is composed of  two runtime environments} respectively for two heterogeneous workloads.
  \item For a runtime environment, if we can not find another coordinated one, we will independently provision that runtime environment.
\end{itemize}

 We call the above model a FLB\_NUB (Fixed Lower Bound and No Upper Bound) model.

\begin{figure}[h]
\centering
\includegraphics[width=2.5in]{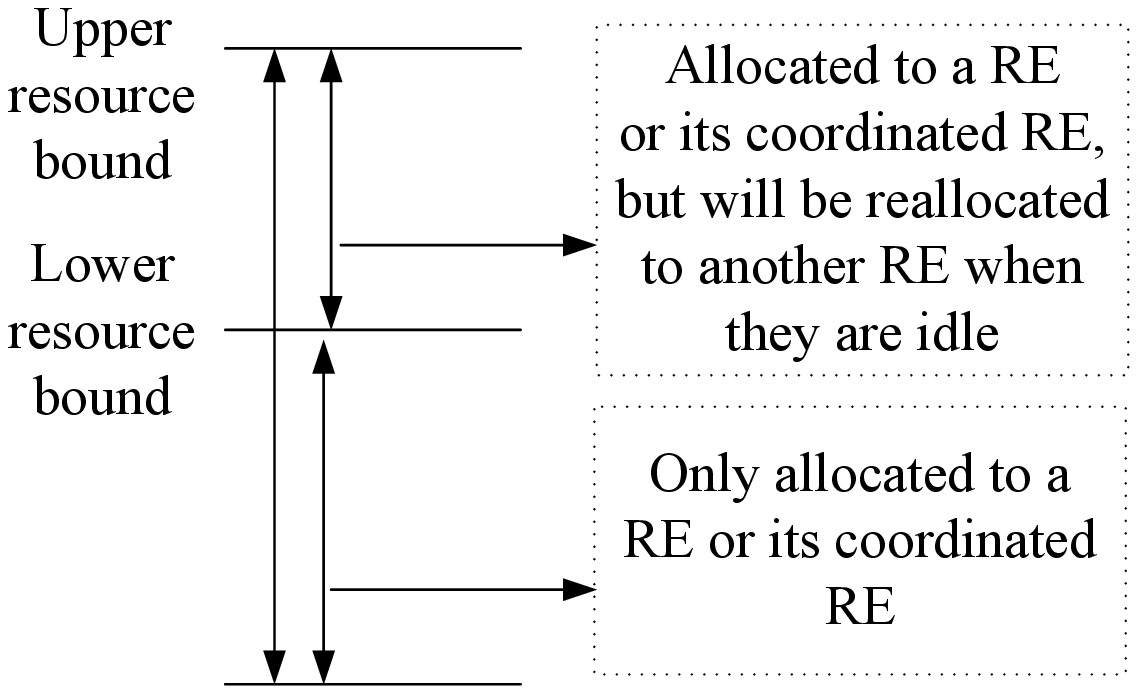}
\caption{Two resource bound sizes. In this figure, RE stands for runtime environment}\label{resource_bound}
\end{figure}

 \item The setup policy.\\
A service provider determines when and how to perform the setup work when resources are dynamically requested or released. The setup work includes provisioning operating systems and configuring applications. For example, if the service provider pays high attention to the security of data, it may require wiping off the operating system and data on disks when a node is released to the resource provider.

\end{enumerate}

\begin{figure}[h]
\centering
\includegraphics[width=2.5in]{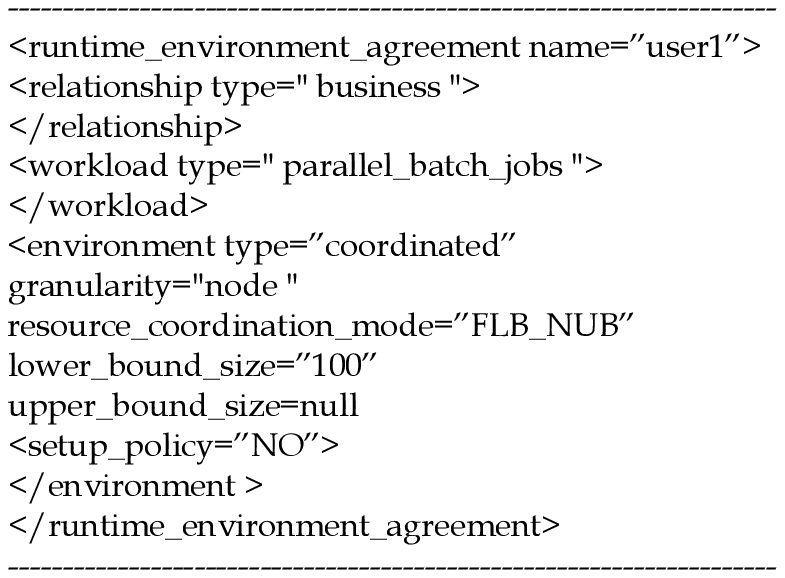}
\caption{A part of a runtime environment specification}\label{RE_example}
\end{figure}

Fig.\ref{RE_example} gives out a part of a runtime environment specification of parallel batch jobs for Case Three in Section \ref{related_work}. Our runtime environment specification is easily extensible, since we choose the XML (eXtensbile Markup Language) language to express it.

\subsection{PhoenixCloud architecture} \label{architecture}

\textbf{Layered architecture}: PhoenixCloud follows a two-layered architecture: one is \emph{the common service framework} (in short CSF) for a resource provider, and another is  \emph{the thin runtime environment software} (in short, TRE) for a service provider. The two-layered architecture has two implications: first, there lies a separation between the CSF and a TRE. The CSF is provided and managed by a resource provider, independent of any TRE. With the support of the CSF, a TRE or two coordinated TREs can be created on demand for a service provider. Second, for heterogeneous workloads, the common sets of functions of runtime environments are delegated to the CSF, while a TRE only implements the core functions for a specific workload.

\begin{figure}[h]
\centering
\includegraphics[width=3.5in]{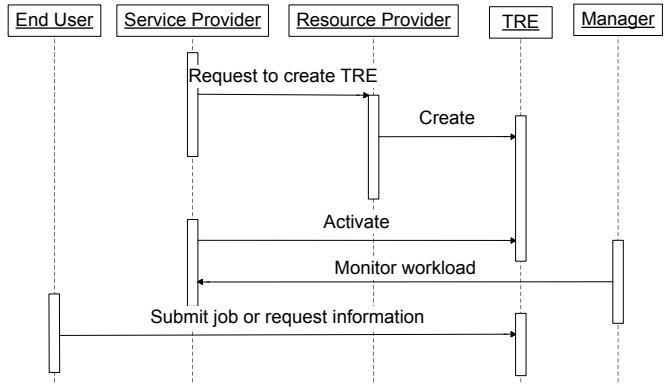}
\caption{Interactions of three user roles in PhoenixCloud}\label{interactions}
\end{figure}

As shown in Fig.\ref{interactions}, there are three interacting user roles in PhoenixCloud: a resource provider, service providers and end users:
 	\begin{itemize}
    \item  The CSF is running on the Cloud site. A resource provider is responsible for provisioning runtime environments with the support of the CSF.
    \item The CSF provides a Web portal for a service provider to describe its runtime environment requirements. After a service provider has requested to create a runtime environment, the CSF is responsible for deploying and starting a TRE.
    \item After a service provider has activated its runtime environment, a service provider has an associated \emph{Manager} that monitors workload changes and resources status. \emph{Manager} is a core component of a TRE. Each \emph{Manager} requests or releases resources on behalf of a service provider according to load status and resources status.
    \item After a runtime environment is providing service, end users use \emph{Web Portal} to submit jobs or send requests.
  \end{itemize}

The advantages of separating the CSF and a TRE are twofold: first, developing a new TRE for different workloads is lightweight, since many common functions have been implemented in the CSF. Secondly, creating a TRE on demand is lightweight, since the CSF is ready and running before any TRE is created.\\

\textbf{Main components of the CSF}: The major components of the CSF are as follows:
\begin{enumerate}
  \item \emph{Lifecycle Management Service} is responsible for managing the lifecycle of a TRE.
  \item \emph{Resource Provision Service} is responsible for provisioning resources to a TRE.
  \item \emph{Virtual Machine Provision Service} is responsible for managing the lifecycle of a virtual machine, such as creating or destroying virtual machine, like XEN.
  \item \emph{Deployment Service} is a collection of services for deploying and booting the operating system, the CSF and TREs. Major services include DHCP, TFTP, and FTP.
  \item \emph{Agent} on each node is responsible for discovering node resources, such as CPU information, memory size and operating system version; downloading required software packages; starting or stopping service daemons, and transferring data.
  \item There are two types of monitors: \emph{Resource Monitor} and \emph{Application Monitor}. \emph{Resource Monitor} on each node monitors usages of physical resources, e.g.  CPU, memory, swap, disk I/O and network I/O; \emph{Application Monitor} monitors application status.
  \item \emph{Process Management Service} is responsible for starting, signaling, killing, and monitoring parallel/sequential jobs.
\end{enumerate}

\textbf{Main components of a TRE}: There are three components in a TRE: \emph{Manager}, \emph{Scheduler} and \emph{Web Portal}. \emph{Manager} is responsible for dealing with users' requests, managing resources and interacting with the CSF. \emph{Scheduler} is responsible for scheduling the user's job or distributing user requests. \emph{Web Portal} is the GUI through which end users submit and monitor jobs or applications. When a TRE is created, a configuration file will describe their dependencies. The details can be found in Section 4 of our previous work \cite{43}.\\

\begin{figure}[h]
\centering
\includegraphics[width=3.5in]{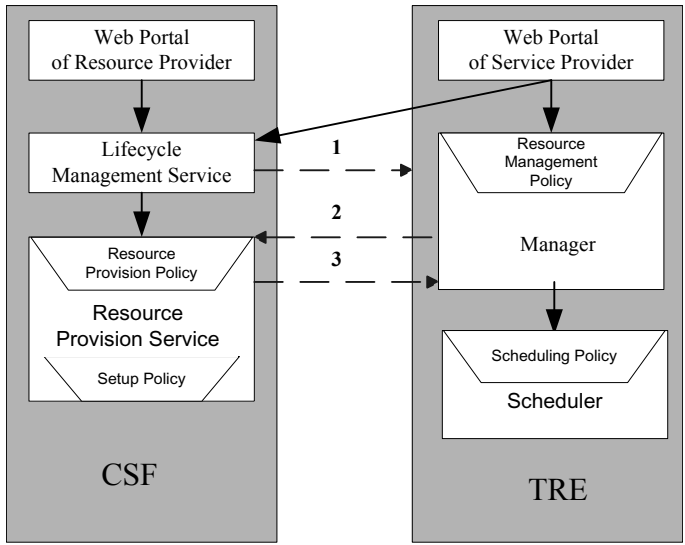}
\caption{The summary of interactions and extension points for management mechanisms of PhoenixCloud. Number 1 indicates creating, destroying, activating and deactivating a TRE; Number 2 indicates requesting and releasing resources; Number 3 indicates proactively provisioning resources.}\label{major_components}
\end{figure}

\textbf{The customized policies of the CSF and a TRE}: Fig.\ref{major_components} shows the major components and their extension points for the management mechanisms.\\
Specified for \emph{Resource Provision Service}, \emph{a resource provision policy} determines when \emph{Resource Provision Service} provisions how many resources to a TRE or how to coordinate resources between two coordinated runtime environments; \emph{a setup policy} determines when and how to do the setup work, such as wiping off the operating system or doing nothing.\\
Specified for \emph{Manager}, \emph{a resource management policy} determines when \emph{Manager} requests or releases how many resources from or to the resource provision service according to what policy.\\
For different workloads, \emph{the scheduling policy} has different implications. For parallel batch jobs, the scheduling policy determines when and how the scheduler chooses parallel jobs for running. For Web service, the scheduling policy includes two policies: \emph{the instance adjustment policy} and \emph{the request distribution policy}. \emph{The instance adjustment policy} decides when the number of Web service instances is adjusted to what an extent, and \emph{the request distribution policy} decides how to distribute requests according to what criteria.

\begin{figure}[h]
\centering
\includegraphics[width=3.5in]{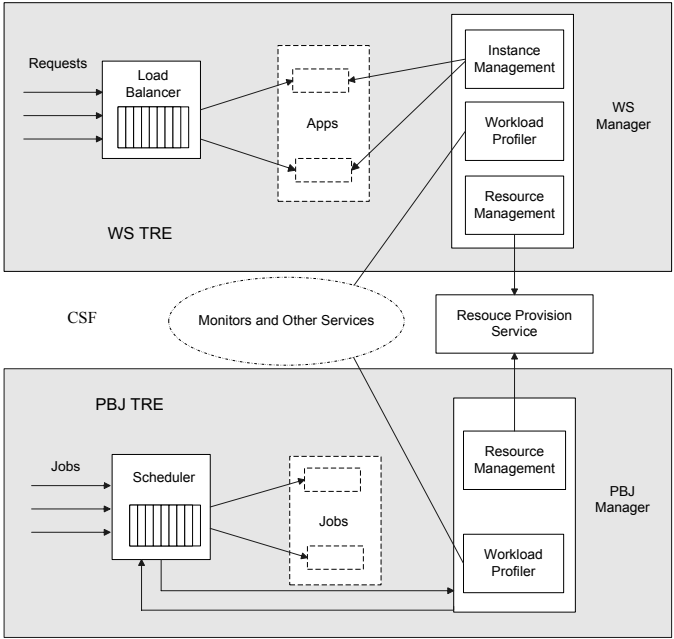}
\caption{Interactions of a PBJ TRE and a WS TRE with the CSF.}\label{CSF_TRE}
\end{figure}

\textbf{Interactions of a TRE with the CSF}:  In the rest of this paper, we call a TRE for parallel batch jobs as PBJ TRE; we call a TRE for Web service as WS TRE. Fig.\ref{CSF_TRE} shows the interactions between TREs and the CSF in two coordinated runtime environments.\\
The interaction of a WS TRE with the CSF is explained as follows:\\
\begin{itemize}
  \item \emph{WS Manager} obtains resources with the size of the lower resource bound from \emph{Resource Provision Service}, and runs Web service instances with a matching scale.\\
  \item \emph{WS Manager} interacts with \emph{Load Balancer} (a type of \emph{Scheduler})to set its request distribution policy. \emph{WS Manager} registers IP and port information of Web service instances to \emph{Load Balancer} that is responsible for assigning workload to Web service instances, and \emph{Load Balancer} distributes requests to Web services instances according to the request distribution policy. We integrate LVS \cite{13} as the IP-level load balancer.\\
  \item \emph{Monitor} on each node periodically checks resources utilization rates and reports to \emph{WS Manager}. If the threshold performance value is exceeded, e.g., the average of utilization rates of CPUs consumed by instances exceeds 80\%, \emph{WS Manager} adjusts the number of Web service instances according to the instance adjustment policy.\\
  \item According to current Web service instances, \emph{WS Manager} requests or releases resources from or to \emph{Resource Provision Service}.\\
\end{itemize}

The interactions of a PBJ TRE with the CSF are explained as follows:\\
\begin{itemize}
  \item  Scheduling events tell \emph{PBJ Manager} to send scheduling commands to \emph{Scheduler}. Scheduling events include the timer registered by the administrator and new job arrivals.\\
  \item \emph{Scheduler} requests jobs and nodes information from \emph{PBJ Manager}, and takes decisions to run jobs according to \emph{a scheduling policy}.\\
  \item  Driven by periodical timers, \emph{PBJ Manager} scans jobs in queue. If the threshold values defined in \emph{a resource management policy} are exceeded, \emph{PBJ Manager} will request or release resources from or to Resource Provision Service.\\
\end{itemize}

\textbf{Lifecycle management of a TRE}: A traditional runtime environment is self-contained. PhoenixCloud facilitates creating a TRE on demand. Each TRE has three states: \emph{uninitialized}, \emph{created} and \emph{running}. The \emph{uninitialized} state indicates the nascent state of a TRE. The \emph{created} state implies that a TRE for the specific workload is configured and deployed on a Cloud site. The \emph{running} state indicates two meanings: first, resources with \emph{the lower bound size} are allocated to a TRE; secondly, a TRE is providing services to end users.\\

By taking the runtime environment specification of Fig.\ref{RE_example} as an example, we introduce the major interactions in the lifecycle management as follows:
\begin{enumerate}
  \item Through \emph{Web Portal} of a resource provider, a service provider creates its account, and then defines its runtime environment requirements
  \item Through \emph{Web Portal} of a resource provider, a service provider sends a message of \emph{creating a runtime environment} to \emph{Lifecycle Management Service}. Then \emph{Lifecycle Management Service} marks the state of the new runtime environment as \emph{uninitialized}.
  \item \emph{Lifecycle Management Service} sends a message of \emph{deploying a runtime environment} to \emph{Agents} on related nodes, which requests \emph{Deployment Service} to download the required software package of the new TRE. After the new TRE is deployed, \emph{Lifecycle Management Service} marks its state as \emph{created}.
  \item A service provider sends a message of \emph{activating a runtime environment} to \emph{Lifecycle Management Service} through \emph{Web Portal} of a resource provider.
  \item \emph{Lifecycle Management Service} sends the configuration information of a new TRE to \emph{Resource Provision Service}, including \emph{the lower resource bound} and \emph{the upper resource bound}, the resource provision model, the setup policy. For a new PBJ TRE, \emph{Resource Provision Service} will search a WS TRE from another service provider for coordinated resource provisioning if a service provider allows it, but does not run another web service workloads.
  \item \emph{Lifecycle Management Service} sends a message of \emph{starting components of the new TRE} to \emph{Agents}. When the components of the new TRE are started, the command parameters will tell the components what policies should be taken. Then \emph{Lifecycle Management Service} marks the state of the new TRE as \emph{running}.
  \item Before resources are provisioned to the new TRE, the setup policy is triggered by \emph{Resource Provision Service}. When the setup work is performed, \emph{Resource Provision Service} notifies \emph{Manager} that resources are ready.
  \item  The new TRE begins providing service to end users through \emph{Web Portal} of the service provider.
  \item  According to load status, \emph{Manager} dynamically requests or releases resources, which will also trigger the setup policy.
\end{enumerate}
To save the space, we omit the processes of deactivating and destroying a TRE.

\textbf{The advantage of PhoenixCloud}:  The advantages of our system are twofold. First, our system facilitates a service provider to express diverse runtime environment requirements, and enables creating runtime environments, especially coordinated runtime environments, on demand. With a runtime environment specification as a basis, our system can adapt to different cases without an architectural change. For example, our system can adapt to three cases in Section \ref{requirement}. Second, our system supports coordinated resource provisioning for heterogeneous workloads, and our experiments in Section \ref{evaluation} will show this benefit.

\section{ RESOURCE COORDINATION AND MANAGEMENT POLICIES} \label{policy}
In this section, we respectively propose policies for FB and FLB-NUB models in consolidating two typical heterogeneous workloads: Web services and parallel batch jobs.

\subsection{The FB policy}
We propose the FB resource coordination policy as follows:
\begin{enumerate}
\item In creating two coordinated runtime environments (a PBJ TRE and a WS TRE) for two heterogeneous workloads, service providers specify the same value for the lower resource bound and the upper resource bound for each runtime environment.
\item \emph{Resource Provision Service} allocates resources with the sizes of the lower resource bounds to two TREs at their startups. \emph{The size of coordinated resources} that are shared by two coordinated runtime environments is the sum of the lower resource bounds of two runtime environments.
\item Resource demands of a WS TRE have high priority than that of a PBJ TRE. If a WS TRE demands resources that can not be satisfied by \emph{Resource Provision Service}, the latter will force the a PBJ TRE to release resources with the size required by a WS TRE, and then reallocate resources to a WS TRE.
\item \emph{Resource Provision Service} registers a periodical timer (\emph{a time unit of leasing resources}) for checking idle resources within the limit of the size of coordinated resources \emph{per time unit of leasing resources}. For example, in EC2, the minimal time unit of leasing resources is one hour. If there are idle resources, \emph{Resource Provision Service} will provision all idle resources to a PBJ TRE.\\

\end{enumerate}

For the above resource provision policy, the matched resource management policy of a PBJ TRE is as follows:
\begin{enumerate}
  \item \emph{PBJ Manager} receives the resources provisioned by \emph{Resource Provision Service}.
  \item If \emph{Resource Provision Service} forces \emph{PBJ Manager} to return resources, the latter will release resources with the required size. If there are no enough idle resources in \emph{PBJ Manager}, it will kill jobs from the beginning of the minimum job size in turn, and then release resources with the required size. If there are more than one running jobs with the same job size, the job with the latest starting time will be killed firstly.
\end{enumerate}

In the rest of this paper, we call the above policies as NLB-NUB policies.
In a recent work of USENIX 09 ATC, W. Zhang et al \cite{42} argue that in managing web services of data centers, actual experiments are cheaper, simpler, and more accurate than models for many management tasks. We also hold the same position. In Section \ref{WorldCup}, we will explain how to obtain the management policies for a specific web service through real experiments.\\
In the rest of this paper, we call the above policies as FB policies.

\subsection{The FLB-NUB policy}
We propose the FLB-NUB resource coordination policy as follows:
\begin{enumerate}
  \item In creating two coordinated runtime environments, service providers only specify the lower resource bound size for each runtime environment with the upper resource bound size undefined.
  \item \emph{Resource Provision Service} allocates resources with the lower bound sizes to a PBJ TRE and a WS TRE at their startups.
  \item \emph{Resource Provision Service} registers a periodical timer (a time unit of leasing resources) for checking idle resources within the limit of the size of coordinated resources per \emph{time unit of leasing resources}. If there are idle resources, \emph{Resource Provision Service} will provision all idle resources to a PBJ TRE.
  \item If a WS TRE demands resources, \emph{Resource Provision Service} will allocate enough resources.
\end{enumerate}

For the above \emph{resource provision policy}, the matched \emph{resource management policy} of a PBJ TRE is as follows:\\
We define \emph{the ratio of adjusting resource} as \emph{the ratio of the accumulated resource demands of all jobs in queue} to \emph{the current resources owned by a TRE}. When the ratio of adjusting resource is greater than one, it indicates that for immediate running, some jobs in the queue need more resources than that currently owned by a TRE.\\
We set two threshold values of adjusting resources, and respectively call them \emph{the threshold ratio of requesting resource} and \emph{the threshold ratio of releasing resource}.\\
The process of requesting and releasing resource are as follows:
\begin{enumerate}
  \item \emph{PBJ Manager} registers a periodical timer (a time unit of leasing resources) for adjusting resources per time unit of leasing resources. Driven by the periodical timer, \emph{PBJ Manager} scans jobs in queue.
  \item If the ratio of adjusting resources exceeds the threshold ratio of requesting resource, \emph{PBJ Manager} will request resources with the size of $DR1$ as follows:\\
$DR1=$(the accumulated resources demand of all jobs in the queue)$-$(the current resources owned by a PBJ TRE)
  \item If the ratio of adjusting resource does not exceed the threshold ratio of requesting resources, but \emph{the ratio of the resource demand of the present biggest job in queue} to \emph{the current resources owned by a TRE} is greater than one, \emph{PBJ Manager} will request resources with the size of $DR2$:\\
$DR2=$ (resources needed by the present biggest job in queue)$-$ (the current idle resources owned by a TRE)\\
When the ratio of \emph{the resources demand of the present biggest job in the queue} to \emph{the current resources owned by a TRE} is greater than one, it implies that the largest job will not run without available resources.
  \item If the ratio of adjusting resources is lower than the threshold ratio of releasing resources, \emph{PBJ Manager} will releases idle resources with the size of $RSS$ (ReleaSing Size).\\
$RSS=$ (the elastic factor) $\times$(idle resources owned by PBJ TRE), where $0 <$ (the elastic factor) $< 1$\\
  \item If \emph{Resource Provision Service} proactively provisions resources to \emph{PBJ Manager}, the latter will receive resources.
\end{enumerate}

\section{PERFORMANCE EVALUATIONS} \label{evaluation}
In this section, for Web services and parallel batch jobs, we compare the performance of PhoenixCloud, DCS and EC2+RightScale.

\subsection{Evaluation metrics} \label{metrics}
For parallel batch jobs, the metrics are as follows:\\
 we choose the well known metrics- \emph{the throughput in terms of the number of completed jobs} \cite{3} \cite{11} to reflect the major concern of a service provider. We use \emph{the average turnaround time per job} to measure the main concern of end users. The average turnaround time of jobs is the time from submitting a job till completing it, averaged over all jobs submitted \cite{11} \cite{20}.\\
For Web service, the metrics are as follows: we choose the well-know metrics, \emph{the throughput in terms of requests per second} to reflect the major concern of a service provider \cite{6} \cite{10}. For end users, we choose \emph{the average response time per requests} to measure the quality of service, which reflects the major concern of end users \cite{6} \cite{10}.\\
For two consolidated workloads, we choose \emph{the total resource consumption} in terms of $node \times hour$ to evaluate the effectiveness of coordinated resource provisioning. We specially care about \emph{the peak resource consumption} that is \emph{the peak value of the resource consumption in terms of nodes}, since it is a key factor in the capacity planning of the system for a resource provider. For the same workload, if the peak resource consumption of a system is higher, the capacity planning of a system is more difficult.\\
We use \emph{the accumulated times of adjusting resources} to evaluate \emph{the management overhead} of a system, since \emph{each event of requesting, releasing or provisioning resources} will trigger a setup action, for example wiping off  the operating system or data. \emph{The accumulated times of adjusting resources} are the times of resources being dynamically requested, released or provisioned when a runtime environment is providing services.\\
All performance metrics are obtained in the same period that is the duration of workload traces.

\subsection{Workload traces} \label{workload_traces}
\begin{enumerate}
  \item The workload traces of parallel batch jobs\\
We choose two typical workload traces from \cite{31}. The utilization rate of all traces in \cite{31} varies from 24.4\% to 86.5\%. We choose one trace with lower load-NASA iPSC trace (46.6\% utilization) and one trace with higher load-SDSC BLUE trace (76.2\% utilization).\\
NASA iPSC is a real trace segment of two weeks from Oct 01 00:00:03 PDT 1993. For NASA iPSC trace, the configuration of the cluster system is 128 nodes. SDSC BLUE is a real trace segment of two weeks from Apr 25 15:00:03 PDT 2000. For SDSC BLUE trace, the cluster configuration is 144 nodes.
  \item 	Web service workload\\
For Web service, we choose a real workload trace, the World Cup workload trace \cite{2} from June 7 to June 20 in 1998. The World Cup workload indeed reflects the nature of a web service workload, of which the ratio of peak load to normal load is high. Through Google scholar search, we can find that this workload is cited frequently (488 cites in June, 3, 2010) in the related research community.
\end{enumerate}

\subsection{Experiment methods} \label{experiment_method}
\begin{figure}[h]
\centering
\includegraphics[width=2in]{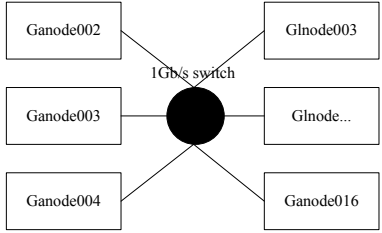}
\caption{The testbed.}\label{testbed}
\end{figure}

To evaluate and compare the dedicated cluster system system, PhoenixCloud, and EC2+RightScale, we adopt the following experiments methods.\\
\begin{itemize}
  \item The real experiments of World Cup workload.\\
  For web service, we obtain \emph{a resource consumption trace} through the real experiments that deploys a WS TRE for the World Cup workload.
  \item The simulated experiments of consolidating two heterogeneous workloads.\\
  The period of a typical workload trace is often weeks, or even months. To evaluate a system, many key factors have effects on experiment results, and we need to frequently perform time consuming experiments. So we use the simulation method to speedup experiments. We speed up the submission and completion of jobs by a factor of 100. This speedup allows two weeks trace to complete in about three hours.\\
  \item The simulated clusters.\\
  The workload traces are obtained from platforms with different configurations. For example, \emph{NASA iPSC} is obtained from the cluster system with each node composed of one CPU; \emph{SDSC BLUE} is obtained from the cluster system with each node composed of eight CPU; The World Cup resource consumption trace is obtained from the four-core Intel(R) Xeon(R) platform; In the rest of experiments, \emph{our simulated cluster system is modeled after the \emph{NASA iPSC} cluster, comprising only single-CPU nodes}. So we divide the workload trace of \emph{SDSC BLUE} by 8.\\
  \item Synthetic heterogeneous workloads.\\
  To the best of our knowledge, the real traces of parallel batch jobs and Web service on the same platform are not available. However, the focus of our paper is to simulate \emph{the case of consolidating two heterogeneous workloads with different peak resource demands on a Cloud site}. So in our experiments, on a basis of workload traces introduced in Section \ref{workload_traces}, we scale two heterogeneous workload traces with different constant factors. We propose a tuple of $(PRC_{PBJ}, PRC_{WS})$ to represent two synthetic heterogeneous workload traces, where $PRC_{PBJ}$ is the peak resource demand of parallel batch job trace and $PRC_{WS}$ is the peak resource demand of Web service trace. For example, a tuple of $(100, 60)$ that is scaled on a basis of \emph{SDSC BLUE} and \emph{World Cup} traces means that we respectively scale \emph{SDSC BLUE} and \emph{World Cup} traces with two different constant factors, and on the same simulated cluster system, the peak resource demand of \emph{SDSC BLUE} and \emph{World Cup} is respectively 100 nodes and 60 nodes.
  \item The testbed.\\
  Shown in Fig.\ref{testbed}, the testbed includes two types of nodes, nodes with the name starting with \emph{glnode} and nodes with the name starting with \emph{ganode}. The nodes of \emph{glnode} have the same configuration, and each node has 2G memory and two CPUs. Each CPU of the node of \emph{glnode} has four cores, Intel(R) Xeon(R) (2.00GHz). The OS is 64-bit Linux with kernel of 2.6.18-xen. The nodes of \emph{ganode} have same configuration, and each node has 1G memory and 2 CPUs, AMD Optero242 (1.6GHz). The OS is 64-bit Linux with kernel version of 2.6.5-7.97-smp. All nodes are connected with a 1 Gb/s switch.
\end{itemize}

\subsection{The real experiments of World Cup workload} \label{WorldCup}
On each node of \emph{glnode}, we deploy eight XEN \cite{39} virtual machines. For each XEN virtual machine, one core and 256M memory is allocated, and the guest operating system is 64-bit CentOS with kernel version of 2.6.18-XEN.\\
On the testbed, we deploy a WS TRE shown in Fig. \ref{CSF_TRE}. In the experiments, \emph{Load Balancer} is LVS \cite{27} with direct route mode \cite{18}.\\

\begin{figure}[h]
\centering
\includegraphics[width=3.5in]{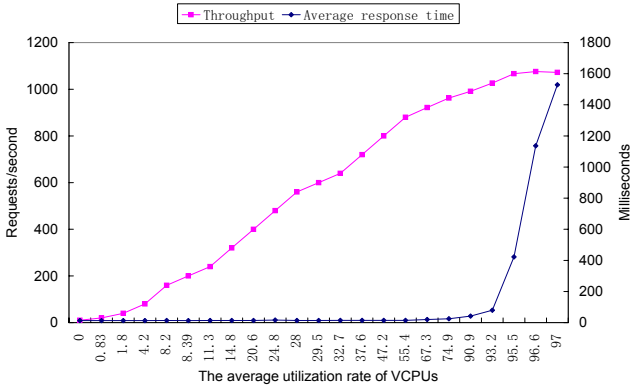}
\caption{Relationship between actual throughput and average utilization rate of VCPUs on the testbed of 16 virtual machines.}\label{throughput_vs_utilization}
\end{figure}

Each \emph{Agent} and \emph{Monitor} are deployed on each virtual machine. LVS and other services are deployed on ganode004, since all of them have light load. We choose the least-connection scheduling policy \cite{18} to distribute requests. We choose httperf \cite{17} as load generator and open source application ZAP! \cite{19} as the target Web service. The versions of httperf, LVS and ZAP! are respectively $0.9.0$, $1.24$ and $1.4.5$. Two httperf instances are deployed on ganode002 and ganode003.\\
The Web workload trace is obtained from the World Cup workload trace \cite{2} with a scaling factor of 2.22. The experiments include two steps. First, we decide the instance adjustment policy; secondly, we obtain the resource consumption trace.\\
In the first step, we deploy PhoenixCloud with the instance adjustment policy disabled. For this configuration, \emph{WS Manager} will not adjust the number of Web service instances. On the testbed of 16 virtual machines, 16 instances of ZAP! are deployed with each instance deployed on each virtual machine. When httperf generates different scale of load, we record the actual throughput, the average response time and the average utilization rate of CPU cores. Since one CPU core is allocated to one virtual machine, for virtual machine, the number of VCPUs is number of CPU cores. So the average utilization rate of each CPU core is also the average utilization rate of VCPUs. Fig.\ref{throughput_vs_utilization} shows the relationship between the actual throughput and average utilization rate of VCPUs. From Fig.\ref{throughput_vs_utilization}, we observe that when the average utilization rate of VCPUs is below 80\%, the average response time of requests is less than $50 milliseconds$. However, when the average utilization rate of VCPUs increases to 97\%, the average response time of requests dramatically increase to $1528 milliseconds$.\\

\begin{figure}[h]
\centering
\includegraphics[width=3.5in]{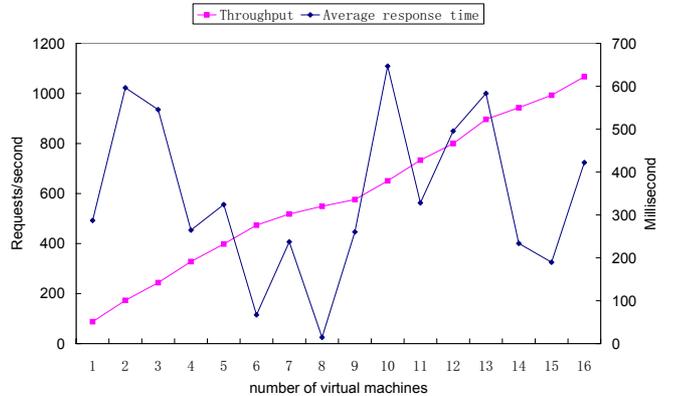}
\caption{Throughput and average response time V.S. number of virtual machines.}\label{throughput_vs_VM}
\end{figure}

\begin{figure}[h]
\centering
\includegraphics[width=3in]{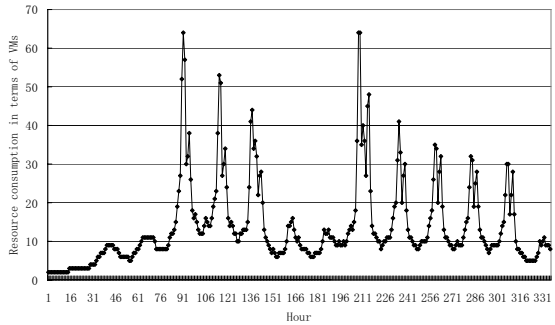}
\caption{The World Cup resource trace in two weeks.}\label{World_resource_trace}
\end{figure}

Based on the above observation, we choose the average utilization rate of VCPUs as the criterion of adjusting the number of instances of ZAP!, and set 80\% as the threshold value.\\
For ZAP!, we specify the instance adjustment policy as follows: the initial number of Web service instances is two. If the average utilization rate of VCPUs consumed by all instances of Web service exceeds 80\% in the past $20 seconds$, \emph{WS Manager} will add one instance. If the average utilization rate of VCPUs, consumed by the current instances of Web service, is lower than $(80\% (\frac{n-1}{n}))$ in the past $20 seconds$, and $n$ is the number of current instances, \emph{WS Manager} will decrease one instance.\\
In the second step, we deploy PhoenixCloud with the above instance adjustment policy enabled. \emph{WS Manager} adjusts the number of Web service instances according to the instance adjustment policy. In the experiments, we also record the relationship between the actual throughput, the average response time and the number of virtual machine.\\
From Fig.\ref{throughput_vs_VM}, we observe that for different number of VMs, the average response time is below 700 milliseconds and the throughput increases linearly when the number of VM increases. This indicates that the instance adjust policy is appropriate, may not optimal.\\
With the above policies, we obtain the resource consumption trace in two weeks. Fig.\ref{World_resource_trace} shows the World Cup resources consumption trace, of which the peak resources demand is $64 VM$.

In the following simulation experiments, if $PRC_{WS}$ is the same in different $(PRC_{PBJ}, PRC_{WS})$ tuples, we use the same World Cup resource trace as the input of Web services in DCS, PhoenixCloud and EC2+RightScale.

\subsection{Simulation Experiments of dedicated cluster system and PhoenixCloud} \label{simulation_DCS_PhoenixCloud}
In this section, we compare DCS and PhoenixCloud in the private Cloud scenario that a resource provider owns the fixed resources that satisfy resource requests of two runtime environments for heterogeneous workloads

\subsubsection{The simulated systems}

\begin{itemize}
  \item The simulated dedicated cluster system \\
  Since the configuration of DCS is decided by the peak resource demand of a workload for a workload tuple $(PRC_{PBJ}, PRC_{WS})$, we presume that the configuration size of the simulated cluster system is the sum of $PRC_{PBJ}$ and $PRC_{WS}$, which is also the smallest valid configuration size. Fig.\ref{simulated_DCS} shows the simulated dedicated cluster system. Resources are statically allocated to two runtime environments: $PRC_{PBJ}$ size for a PBJ RE and $PRC_{WS}$ size for a WS RE. The job simulator is used to simulate the process of submitting job.

\begin{figure}[h]
\centering
\includegraphics[width=3in]{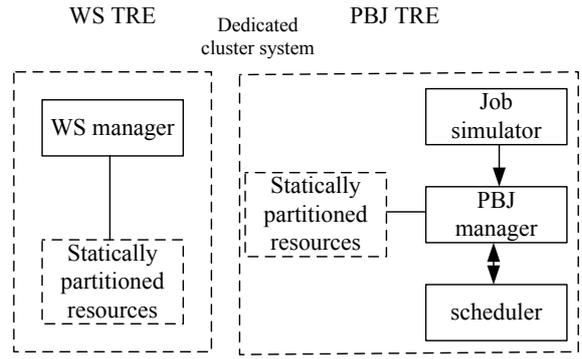}
\caption{The simulated DCS.}\label{simulated_DCS}
\end{figure}

  \item The simulated PhoenixCloud system \\
  For a workload tuple $(PRC_{PBJ}, PRC_{WS})$, in PhoenixCloud, we presume that the bound of the configuration size of the simulated cluster system is the sum of $PRC_{PBJ}$ and $PRC_{WS}$. However, the configuration size of the simulated cluster may decrease to a lower value.\\

\begin{figure}[h]
\centering
\includegraphics[width=3in]{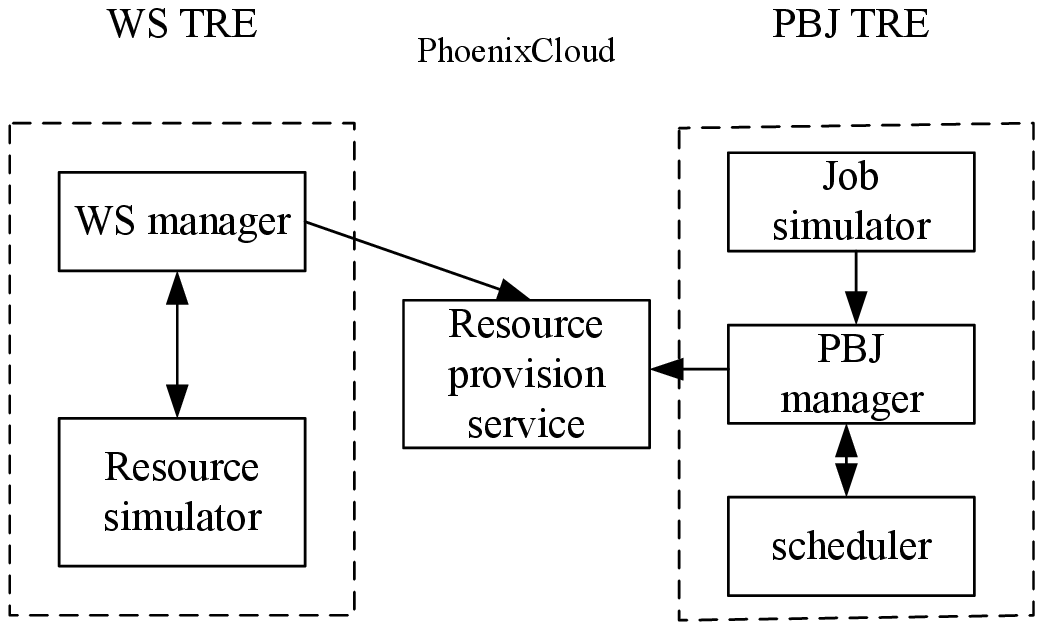}
\caption{The simulated PhoenixCloud system.}\label{simulation_PhoenixCloud}
\end{figure}

In comparison with the real PhoenixCloud system in Fig.\ref{CSF_TRE}, our emulated PhoenixCloud in Fig.\ref{simulation_PhoenixCloud} keeps \emph{Resource Provision Service}, \emph{PBJ Manager}, \emph{WS Manager} and \emph{Scheduler}, while other services are removed. For a WS TRE, the resource simulator simulates the varying resources consumption and drives \emph{WS Manager} to request or release resources from or to \emph{Resource Provision Service}.

\end{itemize}

\subsubsection{Experiment configurations}
\begin{itemize}
  \item The resource coordination and management policy. For DCS, resources are statically allocated to a runtime environment. PhoenixCloud adopts the FB policy.
  \item The scheduling policy. A dedicated cluster system and PhoenixCloud adopt the same first-fit scheduling policy for parallel batch jobs. The first-fit scheduling policy scans all the queued jobs in the order of job arrival and chooses the first job, whose resources requirement can be met by the system, to execute.
\end{itemize}

\subsubsection{Simulation Experiment Results} \label{FB_results}
Table \ref{DCS_PC_IPSC_WCup} and Table \ref{DCS_PC_BLUE_WCup} respectively summarize experiment results for NASA iPSC+World Cup, of which the tuple of peak resource demands $(PRC_{PBJ}, PRC_{WS})$ is $(128, 128)$, and SDSC BLUE+World Cup, of which the tuple of peak resource demands $(PRC_{PBJ}, PRC_{WS})$ is $(144, 128)$. In the rest of this section, we will investigate the effect of varied tuple of peak resource demands.

\begin{table}[h]
\caption{Metrics of DCS and PhoenixCloud for NASA IPSC+WORLD CUP} \label{DCS_PC_IPSC_WCup}
\begin{tabular}{|p{0.8in}|p{0.65in}|p{0.75in}|p{0.6in}|}
\hline
System (configuration size)	&throughput (number of completed jobs)	&average execution time (seconds)	 &average turnaround time  (seconds)\\ \hline
DCS(256)	&2603	&573	&578 \\ \hline
PhoenixCloud (128)	&2549	&520	&839 \\ \hline
PhoenixCloud (152)	&2603	&573	&795 \\ \hline
PhoenixCloud (217)	&2603 	&573	&579 \\ \hline
PhoenixCloud (256)	&2603	&573	&578 \\ \hline
\end{tabular}
\end{table}

\begin{table}[h]
\caption{Metrics of DCS and PhoenixCloud for SDSC BLUE+WORLD CUP.} \label{DCS_PC_BLUE_WCup}
\begin{tabular}{|p{0.8in}|p{0.65in}|p{0.75in}|p{0.6in}|}
\hline
System (configuration size)	&throughput (number of completed jobs)	&average execution time (seconds)	 &average turnaround time  (seconds)\\ \hline
DCS (272)	&2649	&1975	&2667 \\ \hline
PhoenixCloud (144)	&2591	&1983	&7976 \\ \hline
PhoenixCloud (163)	&2648	&1976	&3438\\ \hline
PhoenixCloud (190)	&2652	&1977	&2523\\ \hline
PhoenixCloud (272)	&2657	&1975	&2051\\ \hline
\end{tabular}
\end{table}

From Table \ref{DCS_PC_IPSC_WCup} and Table \ref{DCS_PC_BLUE_WCup}, we can observe two facts: first, using the FB policy in PhoenixCloud, when the configuration size of the simulated cluster is no more than 85\% of that of DCS, the throughput in terms of \emph{the number of completed jobs} of PhoenixCloud is higher than that of DCS (BLUE+WorldCup) or same like that of DCS(iPSC +WorldCup); at the same time, the average turnaround time of PhoenixCloud is better than that of DCS (BLUE+World Cup) or close to that of DCS(iPSC +WorldCup).\\
Second, when the throughput is almost same like that of DCS with a small amount of delay in terms of the average turnaround time (maximally by 38\%), the configuration size of the simulated cluster system can be decreased by about 40\% for two groups of heterogeneous workloads.\\
This is because: (a) for both Web service and parallel batch jobs, ratios of peak loads to normal load are high. However, peak loads of two traces have different timing; (b) when Web service has a short spike, the FB policy will kill running jobs with the smallest resource demands, so we can decrease the configuration size of cluster system, but at the same time increase the average turnaround time.\\
When $PRC_{PBJ}$ is the same, Table \ref{PC_iPSC_WCup} and Table \ref{PC_BLUE_WCup} show the effect of different ratios of $PRC_{WS}$ to $PRC_{PBJ}$ on the performance metrics of PhoenixCloud.

\begin{table}[h]
\caption{Metrics of PhoenixCloud for iPSC+WorldCup.} \label{PC_iPSC_WCup}
\begin{tabular}{|p{0.75in}|p{0.60in}|p{0.50in}|p{0.50in}|p{0.45in}|}
\hline
($PRC_{PBJ}$, $PRC_{WS}$), configuration size &Saved resources(\%) with respect to DCS	 &number of completed jobs	&average turn around time (seconds) & average execution time (seconds)\\ \hline
(128,64),128	&33\%	&2549	&575 &520\\ \hline
(128,128),128 	&50\%	&2549	&839 &520\\ \hline
(128,256),256	&33\%	&2603	&676 &573\\ \hline
\end{tabular}
\end{table}

\begin{table}[h]
\caption{Metrics of PhoenixCloud for BLUE+WorldCup.} \label{PC_BLUE_WCup}
\begin{tabular}{|p{0.75in}|p{0.60in}|p{0.50in}|p{0.50in}|p{0.45in}|}
\hline
($PRC_{PBJ}$, $PRC_{WS}$), configuration size &Saved resources(\%) with respect to DCS	 &number of completed jobs	&average turn around time (seconds) & average execution time (seconds)\\ \hline
(144,64),144	&31\%	&2636	&3343 &1973\\ \hline
(144,128),144 	&47\%	&2591	&7976 &1983\\ \hline
(144,256),256	&36\%	&2657	&2609 &1975\\ \hline
\end{tabular}
\end{table}

From Table \ref{PC_iPSC_WCup} and Table \ref{PC_BLUE_WCup}, we can observe that when two peak resource demands in $(PRC_{PBJ}, PRC_{WS})$ are close, \emph{the percent of saved resources}, which is obtained with the smallest configuration size of cluster, outperforms other cases. This is because when we consolidate two heterogeneous workloads, the configuration size of PhoenixCloud must be greater than the maximum value of two peak resource demands. For parallel batch jobs, if the configuration size of cluster is less than the resource demand of the biggest job, the biggest job can not run. For Web service, if the configuration size of cluster is less than the peak resource demand, overload will happen.

\subsection{ Simulation Experiments of EC2+RightScale and PhoenixCloud} \label{Simulation_EC2_RightScale_PhoenixCloud}
In this section, we compare the performance of PhoenixCloud and EC2+RightScale in the public Cloud scenario. We presume that the simulated cluster system has abundant resources with respect to resource requests of two heterogeneous workloads in both systems.

\subsubsection{ The simulated systems}

\begin{figure}[h]
\centering
\includegraphics[width=3in]{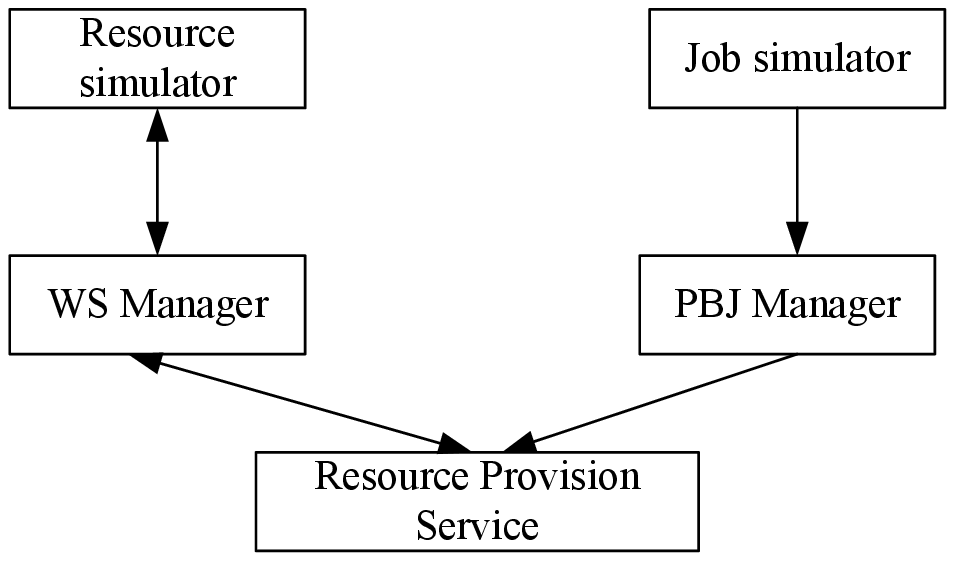}
\caption{The simulated system of EC2+RightScale.}\label{simulation_EC2_RightScale}
\end{figure}

\begin{itemize}
  \item The simulated EC2+RightScale system\\
Because RightScale provides the same scalable management for Web service as PhoenixCloud, we just use the same resource consumption trace for Web service in two systems, which is obtained in Section \ref{WorldCup}. For parallel batch jobs, in EC2, end users simultaneously request resources needed by parallel batch jobs, and the submitted jobs will run immediately, so there is no need for \emph{Scheduler}. Fig. \ref{simulation_EC2_RightScale} shows the simulated architecture of EC2+RightScale.\\
  \item The simulated PhoenixCloud\\
The simulated PhoenixCloud is same as that shown in Fig.\ref{simulation_PhoenixCloud} but with the FLB-NUB policy.\\
\end{itemize}

\subsubsection{ Experiment configurations}

\begin{itemize}
  \item The resource coordination policy. For PhoenixCloud, we adopt the FLB-NUB policy. For EC2+RightScale, there is no resource coordination between two runtime environments.\\
  \item The scheduling policy of parallel batch jobs. PhoenixCloud adopt the first-fit scheduling policy. EC2 needs no scheduling policy, since it is each end user that is responsible for running parallel batch jobs.\\
  \item The resource management policy. For both systems, there is \emph{a time unit of leasing resources}. We presume that \emph{the lease term of a resource is a time unit of leasing resource times a positive integer}. In the EC2+RightScale solution, for parallel batch jobs, each end user is responsible for manually managing resources on EC2 system, and we presume that \emph{a user only releases resources at the end of each time unit of leasing resources if a job runs over}. This is because: a) EC2 charges the usage of resources in terms of \emph{a time unit of leasing resources} (an hour); b) It is difficult for end users to predict the completed time of jobs, and hence releasing resources to resource provider on time is almost impossible. PhoenixCloud adopts the FLB-NUB policy.
\end{itemize}

\subsubsection{ Experiment Results} \label{FLB-results}
Before reporting experiment results, we pick the following parameters as the baseline configuration of PhoenixCloud for comparison, and detailed parameter analysis will be deferred to Section \ref{parameter_analysis}.\\
Through comparisons with a large amount of experiments, we set the baseline parameters in PhoenixCloud:$[B25/U1.2/V0.2/G0.5]$ for iPSC+World Cup and $[B27/U1.2/V0.2/G0.5]$ for SDSC+WorldCup, where $[B25/U1.2/V0.2/G0.5]$ indicates that the size of coordinated resources (which is represented as $B$) is 25 nodes, the threshold ratio of requesting resources (which is represented as $U$) is 1.2; $V0.2$ indicates that the threshold ratio of releasing resources (which is presented as $V$) is 0.2; $G0.5$ indicates that the elastic factor of releasing resources (which is represented as $G$) is 0.5. In both two systems, the time unit of leasing resources (which is represented as $L$) is 60 minute.\\
Table \ref{EC_RightScale_PC_IPSC_WCup} and Table \ref{EC_RIGHTSCALE_PC_BLUE_WCup} respectively summarize the experiment results for iPSC+WorldCup, of which $(PRC_{PBJ}, PRC_{WS})$ is (128, 128), and BLUE+WorldCup traces, of which is $(PRC_{PBJ}, PRC_{WS})$ is (144, 128). \\
From two tables, we can observe two facts:\\
(1) The total resource consumption of PhoenixCloud is less than that of EC2+RightScale (maximally by 28\% and minimally by 14\%) with a small amount of delay in terms of average turnaround time per jobs (maximally by 44\% and minimally by 35\%); \\
(2) PhoenixCloud decreases the peak resource consumption maximally to 31\% with respect to that of EC2+RightScale. This is because PhoenixCloud only requests resources on the condition that the threshold ratio of requesting resources is exceeded, or else jobs will be queued, so PhoenixCloud decreases peak resource consumption and total resource consumption, and increases the average turnaround time.

\begin{table}[h]
\caption{Metrics of EC2+RIGHTSCALE and PhoenixCloud for IPSC +WORLDCUP.} \label{EC_RightScale_PC_IPSC_WCup}
\begin{tabular}{|p{0.45in}|p{0.45in}|p{0.45in}|p{0.45in}|p{0.45in}|p{0.45in}|}
\hline
system	&number of completed jobs	&average turn around time 	&peak resource
consumption 	&total resource consumption &total execution time\\ \hline
EC2 + Right Scale	&2603	&573 seconds	&1319 nodes	&63336 node *hour &573 seconds\\ \hline
Phoenix Cloud 	&2603	&826 seconds	&412 nodes	&45803 node *hour &573 seconds\\ \hline
\end{tabular}
\end{table}

\begin{table}[h]
\caption{Metrics of EC2+RIGHTSCALE and PhoenixCloud for SDSC+WORLDCUP.} \label{EC_RIGHTSCALE_PC_BLUE_WCup}
\begin{tabular}{|p{0.45in}|p{0.45in}|p{0.45in}|p{0.45in}|p{0.45in}|p{0.45in}|}
\hline
system 	&number of completed jobs	&average turn around time 	&peak resource consumption	&total  resource  consumption & total execution time\\ \hline
EC2 + Right Scale	&2657	&1975 seconds	&834 nodes	&45056 node *hour &1975 seconds\\ \hline
Phoenix Cloud 	&2656	&2669 seconds	&468 nodes	&38623 node *hour &1975 seconds\\ \hline
\end{tabular}
\end{table}

When $PRC_{PBJ}$ is the same, Table \ref{PC_iPSC_Ratio} and Table \ref{PC_BLUE_Ratio} show the effect of different ratios of $PRC_{WS}$ to $PRC_{PBJ}$ on the performance metrics of PhoenixCloud. Due to the space limitation, we constrain most of our discussion to the configuration of $BR0.1\_U1.2\_V0.2\_G0.5\_L60$, where $BR0.1$ indicates the ratio of the size of the coordinated resources of PhoenixCloud to the sum of $PRC_{WS}$ and $PRC_{PBJ}$ is 0.1.\\
From Table \ref{PC_iPSC_Ratio} and Table \ref{PC_BLUE_Ratio}, we can observe that when the ratio of $PRC_{WS}$ to $PRC_{PBJ}$ increases, the percent of saved resources (\%) increases, which is obtained against the sum of $PRC_{WS}$ and $PRC_{PBJ}$ times the trace duration. This observation is different from that of the FB policy in Section \ref{FB_results}, this is because in the FLB-NUB policy, resources can be dynamically requested beyond \emph{the lower resource bound}; while in the FB policy, the resources only can be dynamically requested within the limit of \emph{the lower resource bound}.

\begin{table}[h]
\caption{Metrics of PhoenixCloud for iPSC +WorldCup.} \label{PC_iPSC_Ratio}
\begin{tabular}{|p{0.6in}|p{0.65in}|p{0.60in}|p{0.45in}|p{0.55in}|}
\hline
($PRC_{PBJ}$, $PRC_{WS}$ )	&number of completed jobs	&average execution time (seconds)	&average turn around time (seconds)	&saved resources (\%) \\ \hline
(128,64)	&2603	&573	&839	&38.3\% \\ \hline
(128,128)	&2603	&573	&826	&46.8\% \\ \hline
(128,256)	&2603	&573	&839	&58.5\% \\ \hline
\end{tabular}
\end{table}

\begin{table}[h]
\caption{Metrics of PhoenixCloud for BLUE+WorldCUP.} \label{PC_BLUE_Ratio}
\begin{tabular}{|p{0.6in}|p{0.65in}|p{0.60in}|p{0.45in}|p{0.55in}|}
\hline
 ($PRC_{PBJ}$, $PRC_{WS}$ )	&number of completed jobs	&average execution time (seconds)	&average turn
 around time (seconds)	&saved resources (\%) \\ \hline
(144,64)	&2654	&1974	&2682	&52.0\% \\ \hline
(144,128)	&2656	&1975	&2669	&57.7\% \\ \hline
(144,256) 	&2654	&1974	&2761	&64.5\% \\ \hline
\end{tabular}
\end{table}

\subsubsection{ Parameter Analysis} \label{parameter_analysis}
\begin{figure}[ht]
\centering
\subfloat[iPSC+WorldCup]{\includegraphics[width=3.0in]{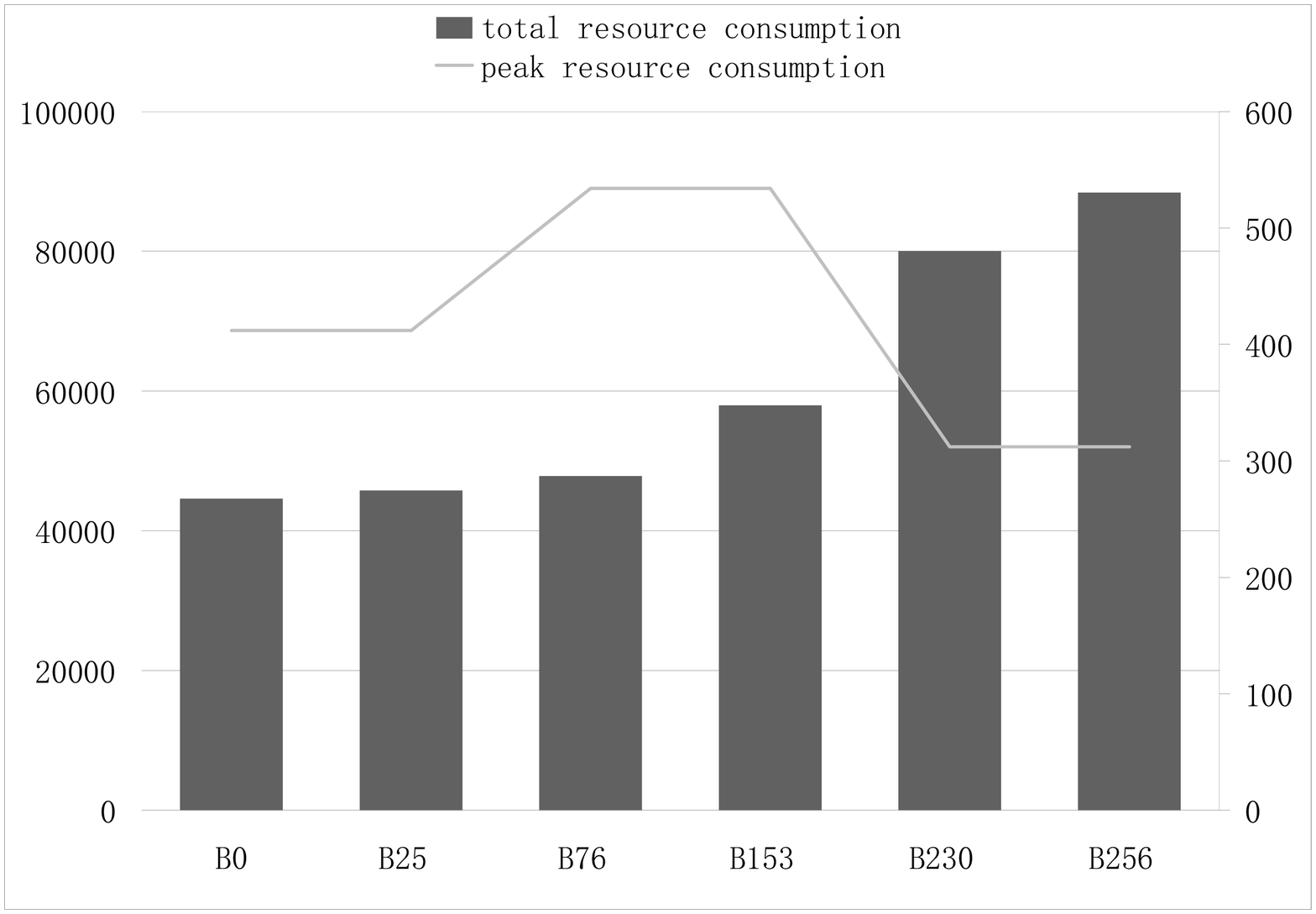}}\\
\subfloat[Blue+WorldCup]{\includegraphics[width=3.0in]{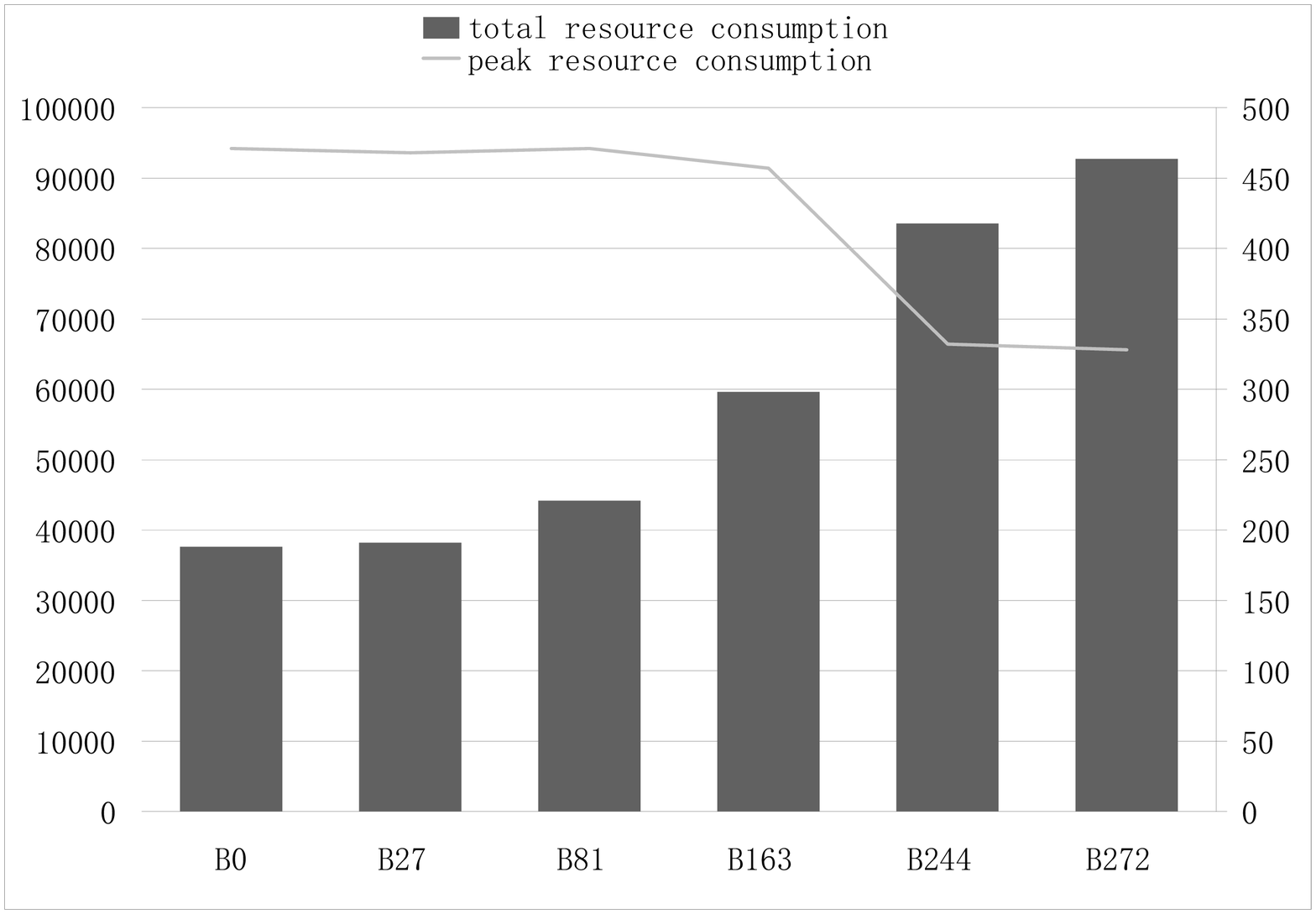}}\\
\caption{Peak and total resource consumptions V.S. different $B$.} \label{peak_total_resource_vs_B}
\end{figure}

\begin{figure}[ht]
\centering
\subfloat[iPSC+WorldCup]{\includegraphics[width=3.0in]{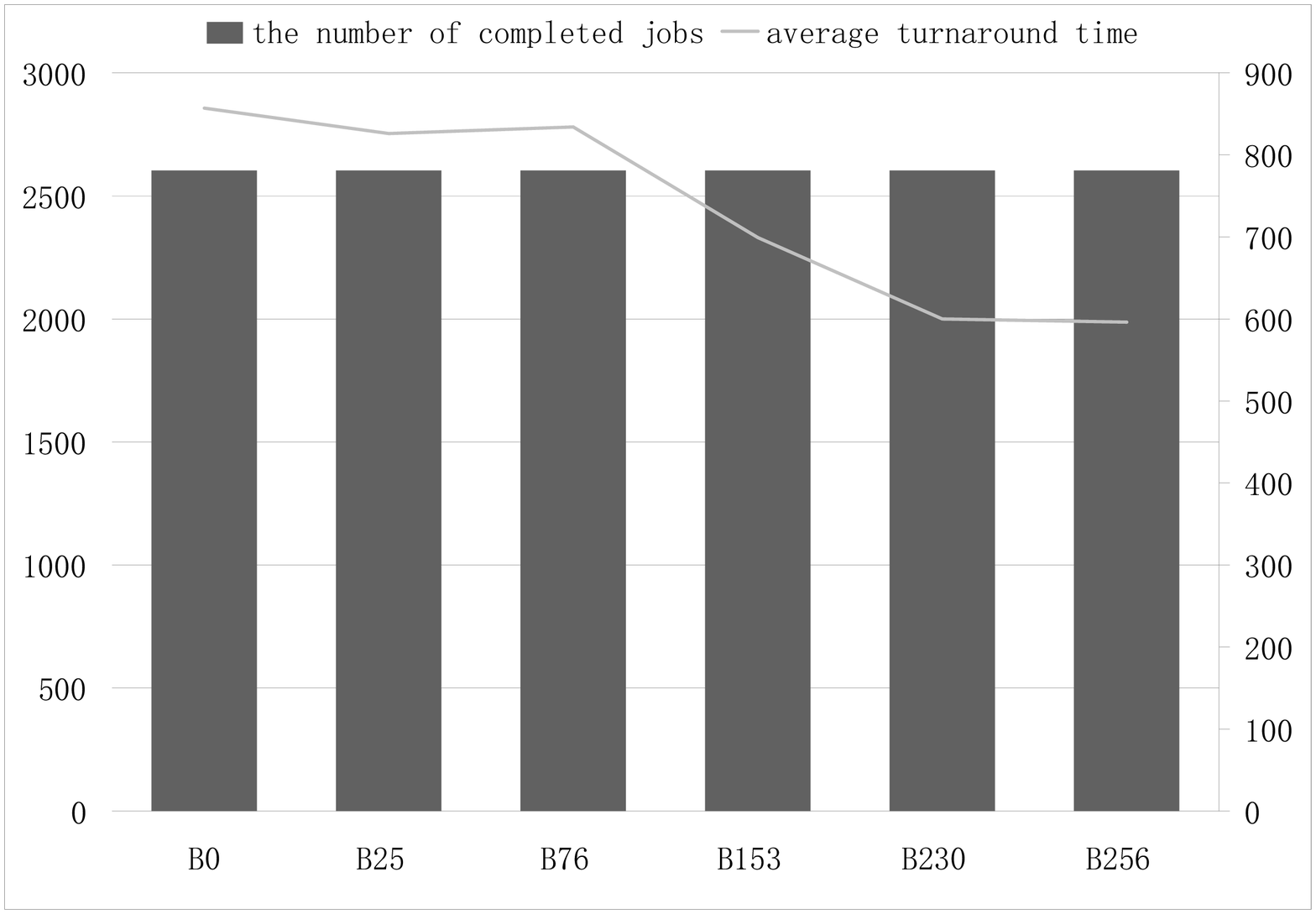}}\\
\subfloat[Blue+WorldCup]{\includegraphics[width=3.0in]{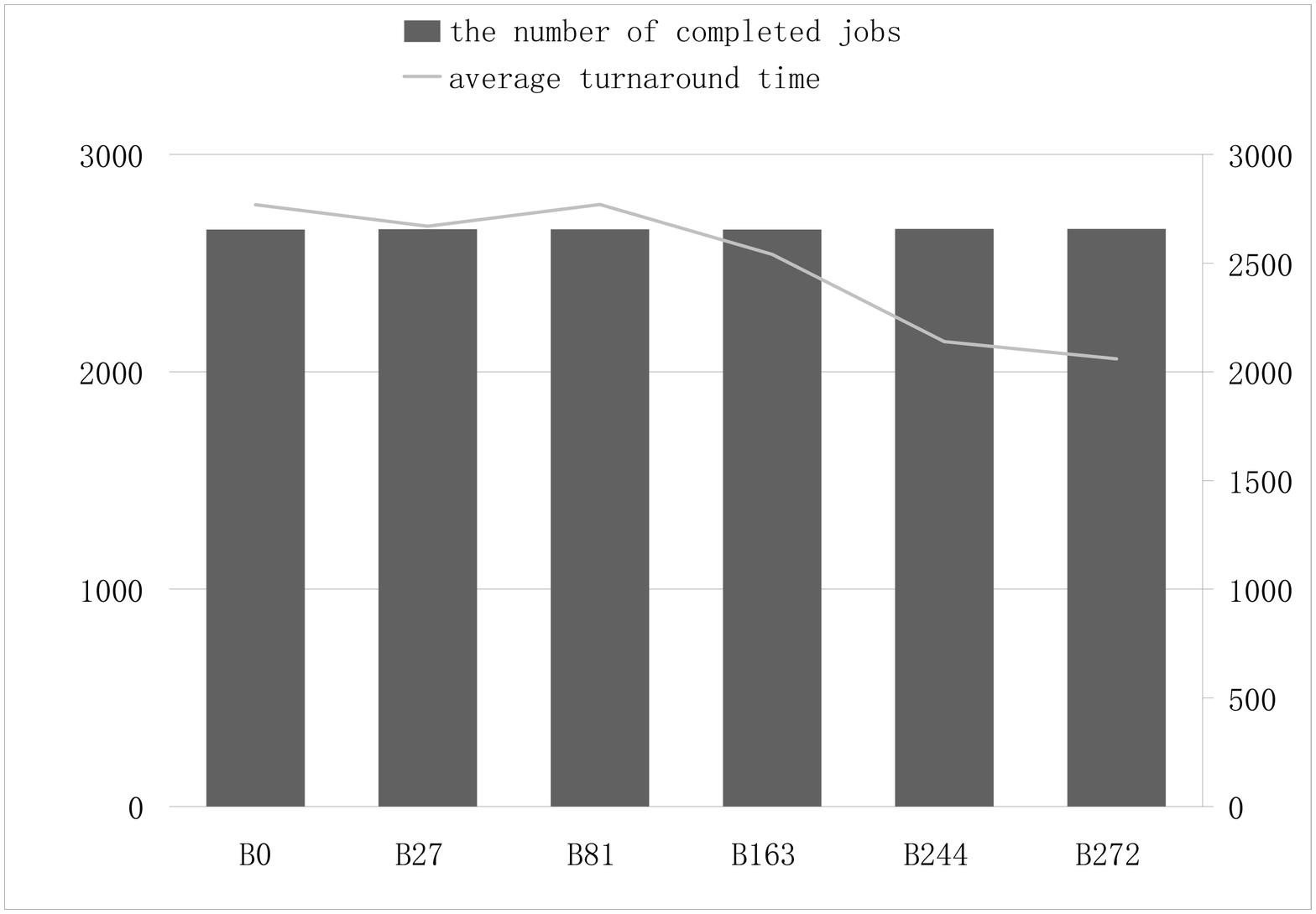}}\\
\caption{The number of completed jobs and average turnaround time V.S. different $B$.} \label{jobs_turnaround_vs_B}
\end{figure}

Because of space limitation, we are unable to present the data for the effect of all parameters; instead, we constrain most of our discussion to the configuration that one parameter varies while the other parameters keep the same values as those of the baseline configuration in Section \ref{FLB-results}, which are representative of the trends that we observe across all cases.\\
\textbf{The effect of the size of coordinated resources ($B$)}. To save space, in PhoenixCloud we tune $B$, while other parameters are $[U1.2/V0.2/G0.5/L60]$. Fig.\ref{peak_total_resource_vs_B} and Fig.\ref{jobs_turnaround_vs_B} show the effect of different $B$ values for two groups of heterogeneous workloads. In the rest of this section, tuples of $(PRC_{PBJ}, PRC_{WS})$ of iPSC+WorldCup and BLUE+WorldCUP are respectively $(128, 128)$ and $(144, 128)$.\\
From Fig.\ref{peak_total_resource_vs_B} and Fig.\ref{jobs_turnaround_vs_B}, we have the following observations:\\
1) With the increase of $B$, the total resource consumption increases, while the average turnaround time decreases. This is because resources under the lower resource bound are only allocated to PBJ TRE and WS TRE, and hence idle resources will also increase when $B$ increases for the same workload; at the same time, with the increase of $B$, more resources will be provisioned to PBJ TRE, so the average turnaround time per jobs decreases.\\
2) The change of $B$ has small effect on the number of completed jobs. This is because PhoenixCloud can dynamically request resources when the threshold ratio of requesting resource is triggered.

\begin{figure}[ht]
\centering
\subfloat[iPSC+WorldCup]{\includegraphics[width=3.0in]{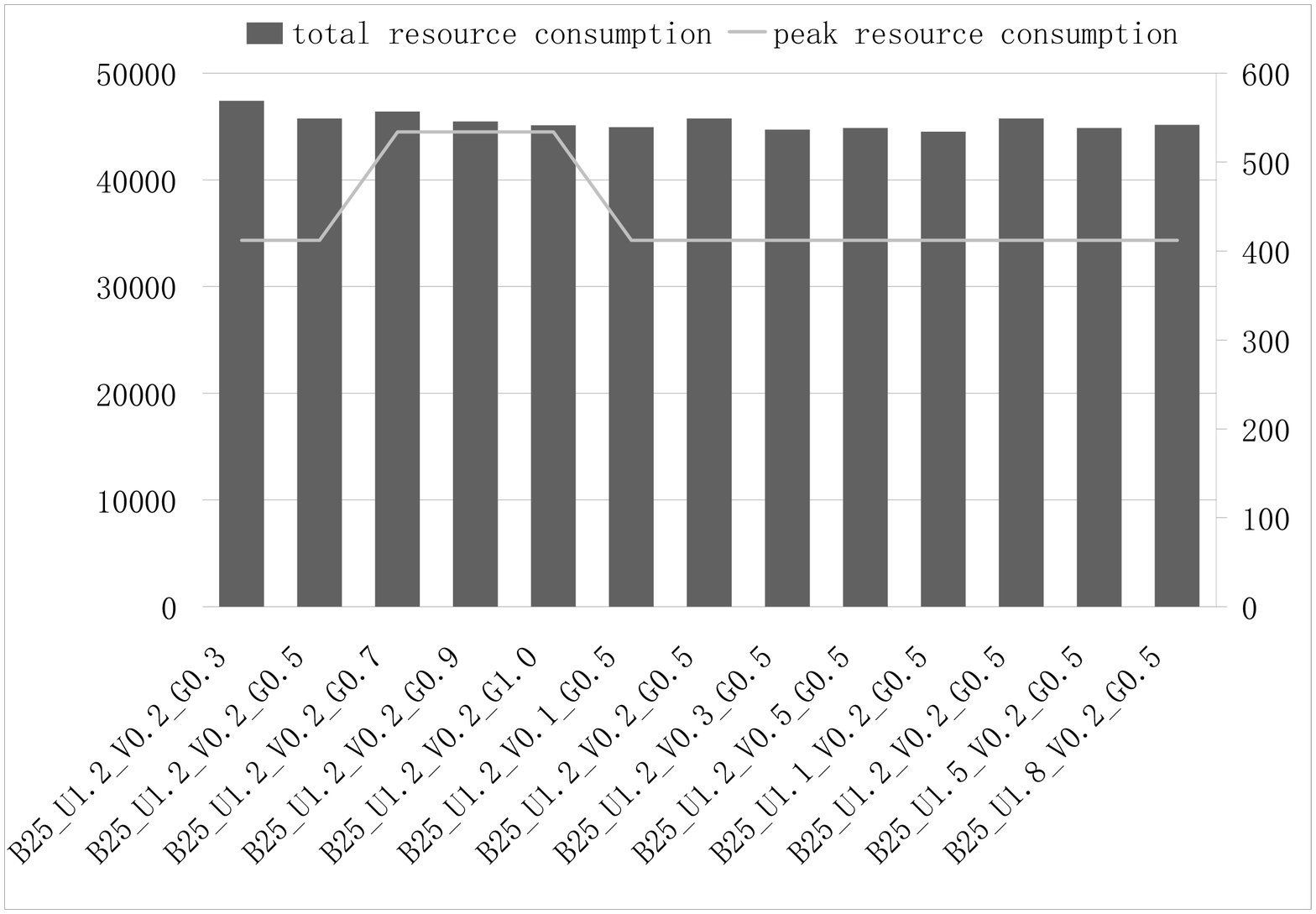}}\\
\subfloat[Blue+WorldCup]{\includegraphics[width=3.0in]{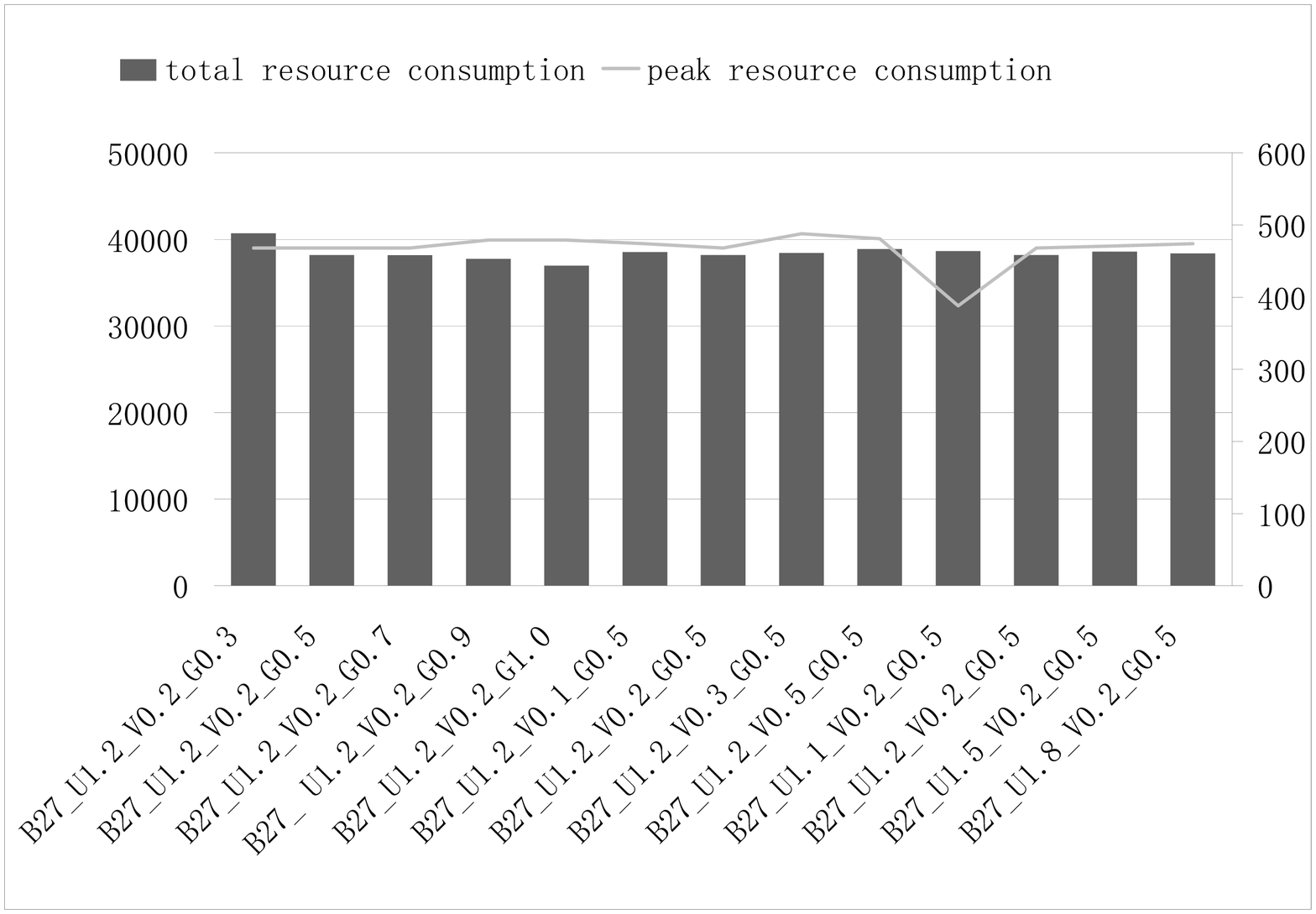}}\\
\caption{peak and total resource consumptions V.S. different $G$, $V$, $U$.} \label{peak_total_resource_vs_GUV}
\end{figure}

\begin{figure}[ht]
\centering
\subfloat[iPSC+WorldCup]{\includegraphics[width=3.0in]{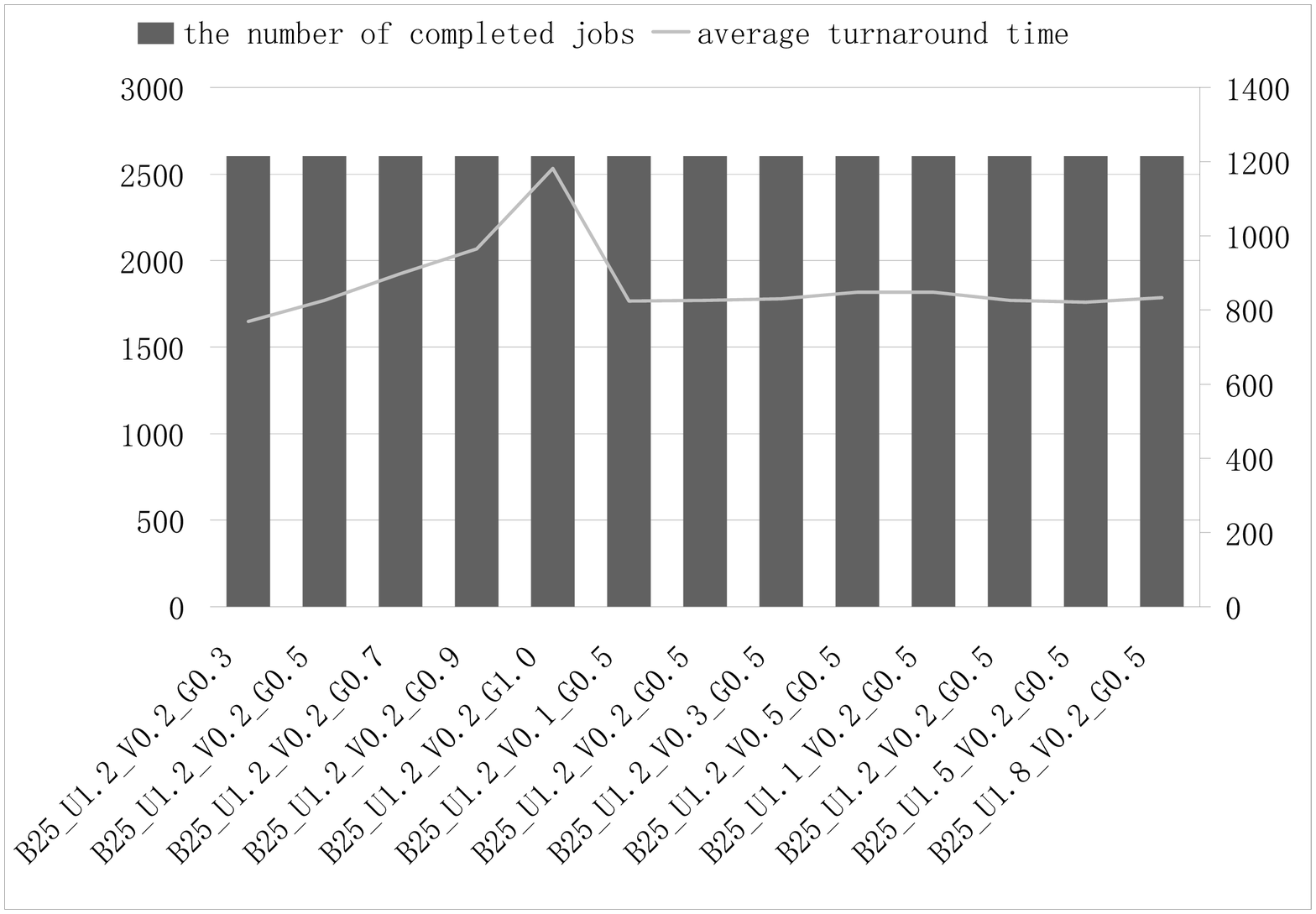}}\\
\subfloat[Blue+WorldCup]{\includegraphics[width=3.0in]{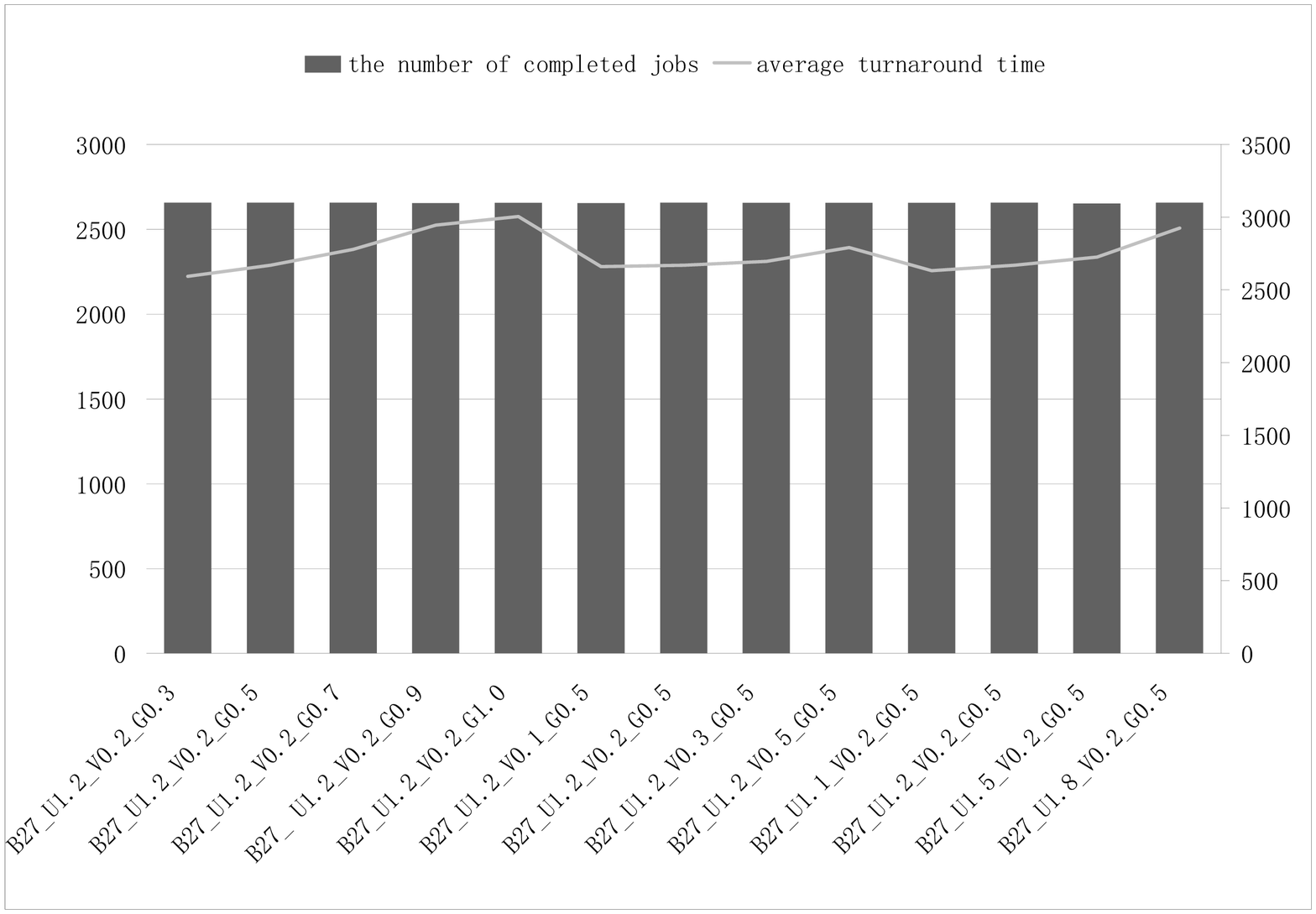}}\\
\caption{the number of completed jobs and average turnaround time V.S. different $G, $V, $U$.} \label{jobs_turnaround_vs_GUV}
\end{figure}

\textbf{The effects of the threshold ratios of requesting resources and releasing resources ($U$ and $V$) and the elastic factor of releasing resource ($G$)}. To save space, in PhoenixCloud we tune one of $U$, $V$, $G$, while other parameters are $[B25/U1.2/V0.2/G0.5/L60]$ for iPSC +WorldCup and $[B27/U1.2/V0.2/G0.5/L60]$for BLUE +WorldCUP. Fig.\ref{peak_total_resource_vs_GUV} and Fig.\ref{jobs_turnaround_vs_GUV} show the effect of different parameters.

\begin{figure}[h]
\centering
\includegraphics[width=3.0in]{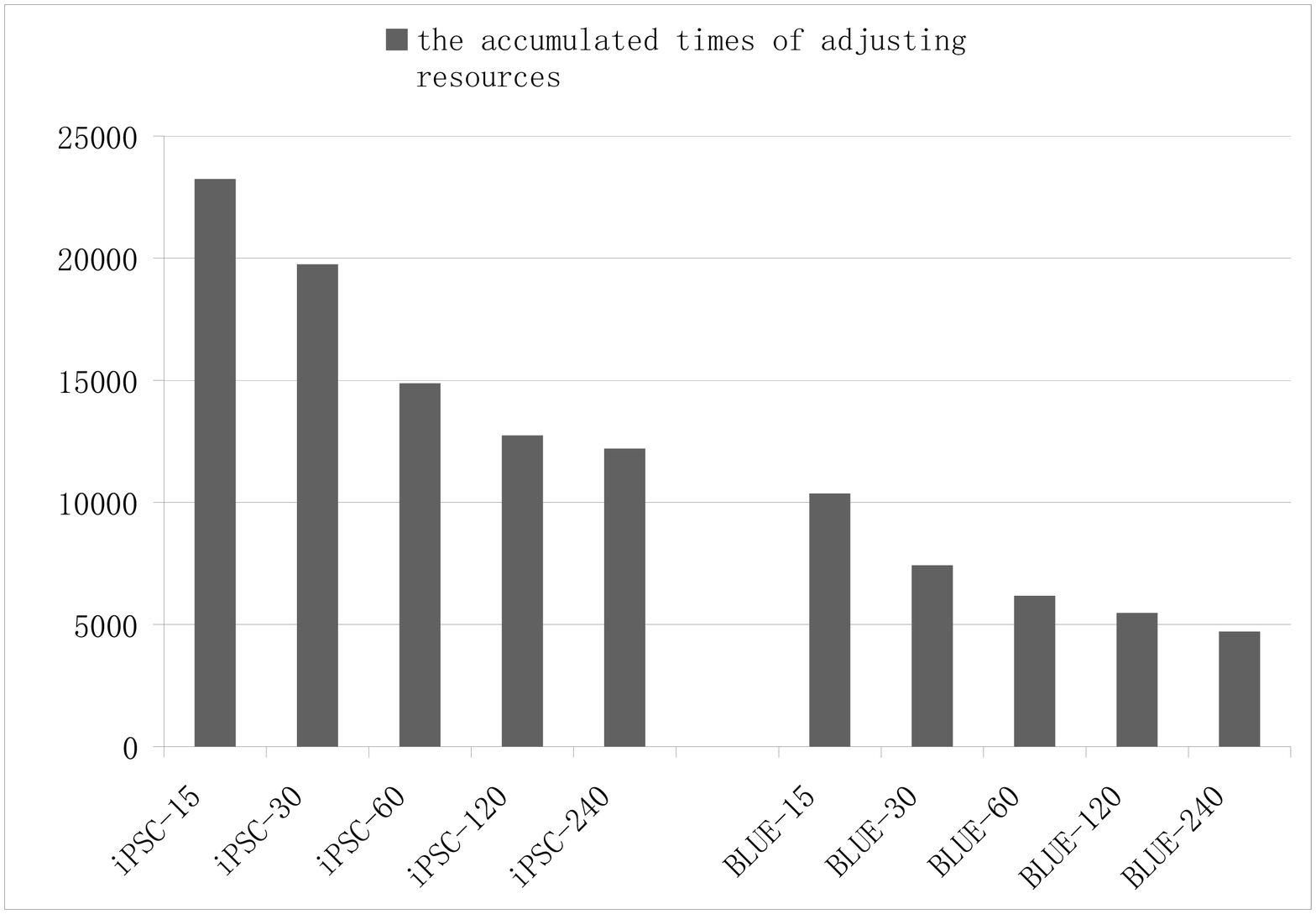}
\caption{management overhead V.S. different time unit of leasing resources.}\label{varied_management_overhead}
\end{figure}

From Fig.\ref{peak_total_resource_vs_GUV} and Fig.\ref{jobs_turnaround_vs_GUV}, we have the following observations:
1) $U$, $V$, $G$ have small effect on the total resource consumption and the number of completed jobs when $B$ is fixed.
2) $G$ is proportional to the average turnaround time when $B$ is fixed.  This is because a larger elastic factor of releasing resources will result in less idle resources when new jobs are submitted. $U$ and $V$ have small effect on the average turnaround time.\\
\textbf{The effect of the time unit of leasing resources}. We respectively set the time unit of leasing resources $L$ as 15/30/60/120/240 minutes, while other parameters are $[B25/U1.2/V0.2/G0.5]$ for NASA iPSC workload and $[B27/U1.2/V0.2/G0.5]$ for SDSC BLUE workload. In Fig.\ref{varied_management_overhead}, iPSC-15 implies that $L$ is 15 minutes and workload is iPSC.

From Fig.\ref{varied_management_overhead}, we have the following observation:\\
1) The management overhead is inversely proportional to $L$. This is because when the time unit of leasing resources is less, the service provider requests resources more frequently.\\
Taking it into account resources are charged at the granularity of a time unit of leasing resources, we make a tradeoff and select $L$ as 60 minutes in PhoenixCloud and EC2+RightScale. In fact, in EC2 system, resources are also charged at the granularity of one hour.\\
\textbf{Implications of Analysis}. Based on the above analysis, we have the following suggestions in choosing factors for two coordinated runtime environments for Web service and parallel batch jobs: since the increase of $B$ value will also result in the increase of total resource consumption, we suggest selecting a low value for $B$ value: about 10\% of the sum of $PRC_{PBJ}$ and $PRC_{WS}$. Increasing the elastic factor of releasing resource $G$ will result in the delay in terms of the average turnaround time. Our experiments show 0.5 makes a good compromise. According to our experiments, when $U$ is greater than 1.0 and less than 2.0, it has a small effect on the metrics in our experiments; when $V$ is greater than 0.1 and less than 0.5, it has a small effect on the metrics in our experiments. So we suggest service providers to choose the baseline configuration in Section \ref{FLB-results} for $U$, $V$, $G$.

\subsection{ Discussions} \label{discussion}
Our experiments show that a service provider has three choices in consolidating heterogeneous workloads:\\
1) If resorting to a private Cloud with the fixed size, it should choose PhoenixCloud with a FB policy. With this solution, the configuration size is smallest with respect to other three solutions. However, this solution increases both the average execution time and the average turnaround time, since jobs may be killed to reallocate resources to web services.

\textbf{In a public Cloud scenario},
2) If paying high attention to the average turnaround time per jobs, it should choose the EC2+RightScale solution. However, this solution will result in higher peak resource consumption, which is several times (two or three in our experiments) of that of PhoenixCloud, and a larger total resource consumption.\\
3) If making a tradeoff among the resource consumption and the average turnaround time of jobs, it should choose PhoenixCloud with the FLB-NUB policy. With this solution, the total and peak resource consumptions of PhoenixCloud are smaller than that of EC2+RightScale, while the average turnaround time is larger than that of EC2+RightScale with small delay.

\section{ CONCLUSIONS} \label{conclusion}
In this paper, we presented a runtime environment specification that expresses diverse runtime environment requirements and built an innovative system PhoenixCloud to enable creating coordinated runtime environments on demand for heterogeneous workloads in different Cloud scenarios. For two typical heterogeneous workloads: Web services and parallel batch jobs, we respectively proposed a coordinated resource provisioning solution in two different Cloud scenarios.\\
For three typical workload traces: SDSC BLUE, NASA iPSC and World Cup, our experiments showed that: a) in the private Cloud scenario, when the throughput is almost same like that of DCS, our solution decreases the configuration size of cluster by about 40\%; b) in the public Cloud scenario, our solution decreases not only the total resource consumption, but also the peak resource consumption maximally to 31\% with respect to that of EC2 + RightScale solution.




\end{document}